 \documentclass{emulateapj}
\usepackage{apjfonts}
\usepackage{natbib}
\usepackage{amssymb, amsmath, amsbsy, epsfig, epsf} %%, slashbox, rotating}
\usepackage[dvips]{color}
\usepackage{float}

\newcommand{\flux}{erg~s$^{-1}$~cm$^{-2}$}
\newcommand{\lum}{erg~s\ensuremath{^{-1}}}
\newcommand{\lbol}{\ensuremath{L\mathrm{_{bol}}}}

\newcommand{\msun}{\ensuremath{M_{\odot}}}
\newcommand{\lsun}{\ensuremath{L_{\odot}}}
\newcommand{\lsunv}{\ensuremath{L_{\odot,V}}}
\newcommand{\kms}{\ensuremath{\mathrm{km~s^{-1}}}}
\newcommand{\ergs}{\ensuremath{\mathrm{erg~s^{-1}}}}  %% not `ergs', but `erg' ---
\newcommand{\mbh}{\ensuremath{M_\mathrm{BH}}}

\newcommand{\pnull}{\ensuremath{P_{\mathrm{null}}}}

\newcommand{\chisq}{\ensuremath{\chi^2}}

\newcommand{\ha}{H\ensuremath{\alpha}}
\newcommand{\hb}{H\ensuremath{\beta}}
\newcommand{\hc}{H\ensuremath{\gamma}}
\newcommand{\hd}{H\ensuremath{\delta}}

\newcommand{\pa}{P\ensuremath{\alpha}}
\newcommand{\pb}{P\ensuremath{\beta}}
\newcommand{\pc}{P\ensuremath{\gamma}}
\newcommand{\pd}{P\ensuremath{\delta}}

\newcommand{\hii}{H\,{\footnotesize II}}
\newcommand{\nii}{[N\,{\footnotesize II}]}
\newcommand{\sii}{[S\,{\footnotesize II}]}
\newcommand{\oiii}{[O\,{\footnotesize III}]}

\newcommand{\feii}{{\rm Fe\,{\footnotesize II}}}

\newcommand{\neii}{Ne\,{\footnotesize II}}
\newcommand{\neiii}{Ne\,{\footnotesize III}}
\newcommand{\nev}{Ne\,{\footnotesize V} }
\newcommand{\oi}{[O\,{\footnotesize I}]}
\newcommand{\oii}{[O\,{\footnotesize II}]}
\newcommand{\ovi}{[O\,{\footnotesize VI}]}
\newcommand{\civ}{C\,{\footnotesize IV}}

\newcommand{\caii}{Ca\,{\footnotesize II}}
\newcommand{\hei}{He\,{\footnotesize I}}
\newcommand{\heiabs}{He\,{\footnotesize I}*}
\newcommand{\nai}{Na\,{\footnotesize I}}

\newcommand{\rev}[1]{{\color{black} #1}}

\newcommand{\thisobj}{J1634$+$2049}
\newcommand{\sersic}{S\'{e}rsic}
\newcommand{\etal}{et~al.}

\slugcomment{}
\shorttitle{A He\,I* BALQSO in the Transitional Phase}
\shortauthors{W.-J.~Liu et~al.}

\begin{document}

%\title{SDSS\,J163459.82$+$204936.0: A Ringed Infrared-Luminous Quasar with Both Absorption Line and Emission 
%  Line Outflows Emerging Out of the Dust-enshrouded Phase}
\title{SDSS\,J163459.82$+$204936.0: A Ringed Infrared-Luminous Quasar with Outflows 
       in Both Absorption and Emission Lines}

  \author{Wen-Juan~Liu\altaffilmark{1,2,3}, Hong-Yan~Zhou\altaffilmark{1,3}, 
          Ning Jiang\altaffilmark{3}, Xufen Wu\altaffilmark{3}, Jianwei Lyu\altaffilmark{4}, 
	  Xiheng Shi\altaffilmark{1}, Xinwen Shu\altaffilmark{5}, Peng Jiang\altaffilmark{1,6,7},  
	  Tuo Ji\altaffilmark{1}, Jian-Guo~Wang\altaffilmark{2},
	  Shu-Fen~Wang\altaffilmark{1,2}, Luming Sun\altaffilmark{1,3}}

\altaffiltext{1}{Polar Research Institute of China, 451 Jinqiao Road, Shanghai 200136, China; 
~ zhouhongyan@pric.gov.cn}
\altaffiltext{2}{Yunnan Observatories, Chinese Academy of Sciences, Kunming, Yunnan 650011, China;
  Key Laboratory for the Structure and Evolution of Celestial Objects, Chinese Academy of Sciences, 
  Kunming, Yunnan 650011, China; wjliu@ynao.ac.cn}
\altaffiltext{3}{Key Laboratory for Research in Galaxies and Cosmology, Department of Astronomy, 
University of Sciences and Technology of China, Hefei, Anhui 230026, China}
\altaffiltext{4}{Steward Observatory, University of Arizona, 933 North Cherry Avenue, Tucson, AZ 85721, USA}
\altaffiltext{5}{Department of Physics, Anhui Normal University, Wuhu, Anhui 241000, China}
\altaffiltext{6}{School of Astronomy and Space Science, Nanjing University, 22 Hankou Road,Nanjing 210093, China}
\altaffiltext{7}{Key Laboratory of Modern Astronomy and Astrophysics (Nanjing University), Ministry of Education, Nanjing 210093, China}
%\altaffiltext{8}{Yunnan Observatories, Chinese Academy of Sciences, Kunming, Yunnan 650011, China}

\begin{abstract}
  SDSS\,J163459.82$+$204936.0 is a local ($z = 0.1293$) infrared-luminous 
  quasar with $L_{\rm IR} = 10^{11.91} \lsun$. 
  We present a detailed multiwavelength study of both the host galaxy and 
  the nucleus. 
  The host galaxy, appearing as an early-type galaxy in the optical images 
  and spectra, demonstrates violent, obscured star formation activities with
  $SFR \approx 140 \msun {\rm yr^{-1}}$, estimated from either the polycyclic 
  aromatic hydrocarbon emission or IR luminosity.
  %% below revised on 2016Feb27: ----
  The optical to NIR spectra exhibit a blueshifted narrow cuspy component
  in \hb, \hei\,$\lambda\lambda$5876,10830 and other emission lines
  consistently with an offset velocity of $\approx900$~\kms,
  as well as additional blueshifting phenomena in high-ionization lines
  (e.g., a blueshifted broad component of \hei$\lambda$10830 and 
  the bulk blueshifting of \oiii$\lambda$5007),
  while there exist blueshifted broad absorption lines (BALs) in \nai~D and 
  \hei$\lambda\lambda3889, 10830$,
  indicative of the active galactic nucleus (AGN) outflows producing BALs and emission lines. 
  %% below is old ---
  %The UV to NIR spectra show both narrow and broad blueshifted emission lines, 
  %as well as blueshifted broad absorption lines (BALs) in \nai~D and 
  %\hei$\lambda\lambda3889, 10830$,
  %indicative of the AGN outflows producing BALs and emission lines. 
  %% ----
  Constrained mutually by the several BALs in the photoionization simulations
  with {\it Cloudy}, the physical properties of the absorption line outflow are
  derived as follows: density $10^{4} < n_{\rm H} \lesssim 10^5$ cm$^{-3}$,
  ionization parameter $10^{-1.3} \lesssim U \lesssim 10^{-0.7}$ and column 
  density $ 10^{22.5} \lesssim N_{\rm H} \lesssim 10^{22.9}$ cm$^{-2}$,
  which are similar to those derived for the emission line outflows.
  This similarity suggests a common origin.
  %% below is waste ---
  %The ionic column densities measured from the absorption troughs and 
  %the line ratios of the blueshifted emission lines allow us to derive a photoionization
  %solution to constrain well the outflowing gas.  %%-- repeated with the detailed sentences later.
  %
  %The similar physical properties of the absorption line outflow and emission
  %line outflow imply a common origin of outflowing gas.
  %% -- first: to say both are of the same origin; then: to say taking advantages of both....
  %%
  Taking advantages of both the absorption lines and outflowing emission lines, 
  we find that the outflow gas is located at a distance of $\sim$ 48\,--\,65~pc from the 
  nucleus, and that the \rev{kinetic luminosity} of the outflow is \rev{$10^{44}$--$10^{46}$ \ergs}. 
  %accounting for more than 10\% of the bolometric luminosity of this object.
  \thisobj\ has a off-centered galactic ring on the scale of $\sim 30$~kpc that is
  proved to be formed by a recent head-on collision by a nearby galaxy for which we 
  spectroscopically measure the redshift.
  Thus, this quasar is a valuable object in the transitional phase emerging out of dust 
  enshrouding as depicted by the co-evolution scenario invoking galaxy merger 
  (or violent interaction) and quasar feedback.
  Its proximity enables our further observational investigations in detail (or tests) 
  of the co-evolution paradigm.
  
\end{abstract}


\setcounter{footnote}{0}
\setcounter{section}{0}

\section{Introduction}

  In the generally believed cold dark matter (CDM) paradigm of the universe, 
  galaxies grow in a ``bottom-up'' fashion as led by the CDM halos, 
  with smaller ones forming first and then merging into successively larger ones.
  Mergers and strong interactions of gas-rich galaxies are also believed to 
  play a vital role in triggering the accretion activity, namely the active 
  galactic nucleus (AGN) phenomenon, of supermassive black holes (SMBHs)
  which reside at the centers of most (if not all) galaxies 
  \citep[see, e.g.,][]{2006ApJS..163....1H}.
  Observationally, the ultraluminous infrared galaxies (ULIRGs; 
  $L_{\rm IR} \geq~10^{12} L_{\sun}$) in the local Universe that were discovered three 
  decades ago are found mostly in mergers,
  which inspired a merger-driven, evolutionary sequence from ULIRGs to quasars 
  and finally to present-day elliptical galaxies 
  \citep{1988ApJ...325...74S,1996ARA&A..34..749S,2006ApJS..163....1H,2008ApJS..175..356H}.
  %%--beginning of the brief description of the scenario--
  At first, galaxy merging induces enormous starbursts, which are almost completely
  enshrouded by dust (i.e., in the ULIRG phase), and triggers the central AGN;
  with the increasing feedback from the starbursts and AGN, the cold gas and
  dust are heated up and even expelled out of the galaxy 
  and thus the AGN becomes optically bright (i.e., the quasar phase);
  meanwhile the large-scale starbursts decline.
  Finally the cold gas and dust is gone, the AGN shuts down, and the
  galaxy becomes an old elliptical.  %%--end of the brief description of the scenario--%%
  This scenario has been supported by subsequent observations and
  $N$-body/SPH simulations \citep{2006ApJS..163....1H}.
  Particularly, the tight correlation between the masses of SMBHs and the 
  properties of the spheroids observed in local quiescent galaxies 
  %host-galaxy spheroids observed in local inactive massive galaxies as an end outcome, 
  suggests a co-evolution of galaxies with SMBHs 
  (see \citealt{2013ARA&A..51..511K} for a review).
  
  %% a new paragraph: the Problems ----
  In practice, however, the concrete triggering and feedback processes 
  underlying this scenario have remained unknown for the last decades.
  This leaves many open questions, e.g., what the timing is between 
  starburst and AGN activities, how the AGN feedback operates in the 
  host galaxy.
  The physical processes actually cannot be learned from the statistical 
  studies alone (e.g., correlation analysis) of galaxy and quasar samples;
  they are also beyond the capabilities  of current simulations
  \citep[see, e.g.,][]{2006ApJS..163....1H,2009ApJS..182..628V}.
  A straightforward way is to carry out complementary investigations of 
  individual sources in detail, particularly of the rare cases in 
  transitional phases of the proposed evolutionary sequence.

  %% J1634 comes to the stage here----
  This paper presents a detailed multiwavelength analysis of 
  SDSS~J163459.82$+$204936.0 (hereafter \thisobj), a type-1 AGN at $z=0.1293$ 
  with outflows revealed in both broad absorption lines (BALs) and narrow emission 
  lines.
  This object was detected by the {\it Infrared Astronomical Satellite (IRAS)} and was 
  compiled by \citet{1995AJ....109.2318C} into their catalog of radio-detected bright 
  {\it IRAS} sources.
  \thisobj\ was noted by us from the SDSS spectral data set when we compiled the 
  sample of low-$z$ quasars with broad \hei$\lambda3889$ absorption troughs 
  \citep{2015ApJS..217...11L}.
  In the present paper, we will see that \thisobj\ is a LIRG with a total infrared 
  luminosity of $10^{11.91}~\lsun$, suggesting a 
  strong ongoing star formation ( SFR $\sim~140~\msun$ ~yr$^{-1}$) .
  Both star formation regions and the AGN show considerable internal dust obscuration.
  The spectroscopy observation on the nearby galaxy demonstrates that \thisobj\ was collided
  through by a galaxy, leaving a stellar ring around it on scales of 30~kpc (\S\ref{subsec:ring}).
  Besides the outflow revealed in BALs of \heiabs\ and \nai~D(\S\ref{subsec:measureabs}), 
  there are outflows revealed in the emission of the narrow Hydrogen Balmer and Paschen lines, 
  \oii, \oiii$\lambda5007$, \hei$\lambda5876$ and $\lambda10830$ (\S\ref{subsec:emissionline}).
  Analyses with model calculations indicate that the physical conditions of the absorption line outflow
  and emission line outflow are similar, suggesting that the two are intrinsically the same outflow.
  The outflow is estimated to be $\sim48$--65~pc away from the central nuclei with
  a large kinetic luminosity $\sim 10^{44}$ \ergs\ (see \S\ref{sec:outflow}).
  In terms of the mid-infrared (MIR) -- far-infrared (FIR) spectral energy distribution (SED) 
  (see \S\ref{subsec:sed}), \thisobj\ is between the prototypal ULIRG/quasar composite object Mrk~231
  \citep[see e.g.,][]{2006ApJ...637..104K, 2013ApJ...764...15V, 2013ApJ...775..127S,2014ApJ...788..123L}
  and normal quasars.
  Mrk~231 has been long-known as the nearest ULIRG/QSO composite object, with a total infrared 
  luminosity of $3.6\times10^{12}L_{\sun}$, AGN bolometric luminosity of $1.5\times10^{46}$~\ergs\ 
  and a star formation rate (SFR) of 170 $\msun \rm yr^{-1}$ \citep{2013ApJ...764...15V}.
  It is a FeLoBAL quasar and displays neutral and ionized nuclear outflows in several optical
  and UV tracers \citep[e.g.,][]{2013ApJ...764...15V,2014ApJ...788..123L};
  recently Mrk~231 has been regarded as the archetype showing galactic-scale quasar-driven winds.
  \thisobj\ should be a young, transitional quasar immediately after Mrk~231
  in the evolutionary sequence, blowing out of the enshrouded dust after a violent
  galactic collision.
  Throughout this work we assume a cosmology with $H_{0} =70$ km~s$^{-1}$~Mpc$^{-1}$,
  $\Omega_{m} = 0.3$, and $\Omega_{\Lambda} = 0.7$.

\section{Observations and Data Analysis}

\subsection{Spectroscopic and photometric Observations}
   
   \thisobj\ has been observed in multiple bands both spectroscopically and
   photometrically.
   It was first spectroscopically observed by SDSS on 2004 August 7th 
   UT with an exposure time of 3072~s under the seeing of $\sim 1\arcsec.3$,
   with a wavelength coverage of 3800--9200~\AA.
   The SDSS pipeline gave a redshift of 0.1286$\pm$0.0014.
   We measure a redshift $z_{\rm em} = 0.1293\pm0.0007$ from \sii$\lambda\lambda6716,6731$,
   and all the following rest frame spectra are referred to this redshift.
   %which we bring the spectra to the rest frame.
   %This object was noted by us as a low-$z$ He\,I* $\lambda3889$ BAL quasar 
   %candidate found in our search of He\,I*$\lambda3889$ BAL quasars in the 
   %SDSS spectroscopic data set \citep{2015ApJS..217...11L}.

   Trying to extend the optical spectrum toward both the near-ultraviolet
   and near-infrared ends, we have taken spectroscopy with the {\it Double Spectrograph} (DBSP)
   on the Palomar 5 m $Hale$ telescope \citep{1982PASP...94..586O}.
   Two exposures of 300s each were obtained on 2014 April 23 UT, when the sky is basically
   clear and the seeing was $\sim 1\arcsec.5$.
   With a 1\arcsec.5 slit-width and the 600/4000 grating,
   the blue side (3150--5700 \AA) spectrum has a spectral 
   resolution of $\sim 4.13$ \AA;
   with a 1\arcsec.5 slit-width and the 600/10000 grating,
   the red-side (7800--10200 \AA) spectrum has a spectral 
   resolution of $\sim 4.2$ \AA.
   The data reduction was performed with the standard routines in the IRAF.\footnote{IRAF is distributed 
   by the National Optical Astronomy Observatories, which are operated by the Association 
   of Universities for Research in Astronomy, Inc., under cooperative agreement with
   the National Science Foundation.}

   In addition, there are two small nearby galaxies seen to the west of \thisobj\ in the
   %%% NOT "on" the West side of ..., but `to the west ...' %%%
   SDSS image (C1, C2 in Figure \ref{fig:2dimage}). The two galaxies show similar
   %%% `we note ...' means: we ask you to notice the following facts...
   colors to \thisobj\ and their photometric redshift values given by SDSS are 0.224$\pm$0.0421
   and 0.283$\pm$0.0781 respectively, close to that of \thisobj\ in light of
   the large uncertainties of the photometric redshifts.
   Considering the galactic ring around \thisobj, we wonder if the two nearby galaxies had been once interacting
   with \thisobj.
   To measure the redshifts of two possible physically companion galaxies, we performed spectroscopy
   observations of \thisobj\ and the two nearby galaxies using the {\it Yunnan Faint Object
   Spectrograph and Camera} (YFOSC) mounted on the Lijiang GMG 2.4m telescope on 2015 March~13.
   The G10 (150 mm$^{-1}$) grating provides a wavelength range of 3400--10000~\AA\
   and a resolution of $R \approx$ 760.
   The 1.\arcsec8 slit-width was adopted, and the slit was rotated
   by a position angle ${\rm PA}=86$\arcdeg\ to place \thisobj\ and C1 and C2
   (see Figure~\ref{fig:companiongal}) all in the slit.
   Two exposures of 2400~s each were obtained. 
   The data reduction was performed with the standard IRAF routines.
   %% 2015June18night, end here---   
   %% 2016Feb27: --- 
   Due to the low spectral resolution and the imperfect HeNeAr lamp spectra,
    IRAF gives a quite large uncertainty of the wavelength calibration, 2.84\AA\ (rms).

   The near-infrared (NIR) spectroscopic observations for this object were performed
   with the TripleSpec spectrograph on the 
   Palomar 5 m Hale telescope  on 2012 April 15. 
   Four exposures of 120~s each were obtained in an A-B-B-A dithering mode, and the sky was clear
   with seeing $\sim$ 1\arcsec.2.
   The slit-width of TripleSpec was fixed to 1\arcsec.
   Two telluric standard stars were taken quasi-simultaneously.
   The data was reduced with the IDL program SpexTool \citep{2004PASP..116..362C}.
   The flux calibration and telluric correction were performed with the IDL program
   using the methods described in \citet{2003PASP..115..389V}.

   MIR Observation of \thisobj\ was performed using the Infrared Spectrograph 
   \citep[IRS;][]{2004ApJS..154...18H} on board $Spitzer$ \citep{2004ApJS..154....1W} 
   on 2008 April 30 (PI: Lei Hao, program ID: 40991).
   All four modes--short-low 1 (SL1), short-low 2 (SL2), long-low 1 (LL1),
   and long-low 2 (LL2)--were used, to obtain full 5--35 $\micron$ low-resolution
   ($R \sim$ 100) spectra. We obtain the reduced spectrum from the public archive
   ``Cornell Atlas of $Spitzer$/IRS Sources''%--
   \footnote{The Cornell Atlas of Spitzer/IRS Sources (CASSIS) is a product of the
    Infrared Science Center at Cornell University, supported by NASA and JPL.
    http://cassis.astro.cornell.edu/atlas/}
    (CASSIS~v7; \citealt{2011ApJS..196....8L}).

   \thisobj\ has been photometrically observed in multiple bands and 
   we list all the available photometric data in Table~\ref{tab:photometry}.

%% $2.2 ----
\subsection{Broadband SED \label{subsec:sed}}

\begin{figure}[htbp] 
      \center
     \includegraphics[width=6.4in]{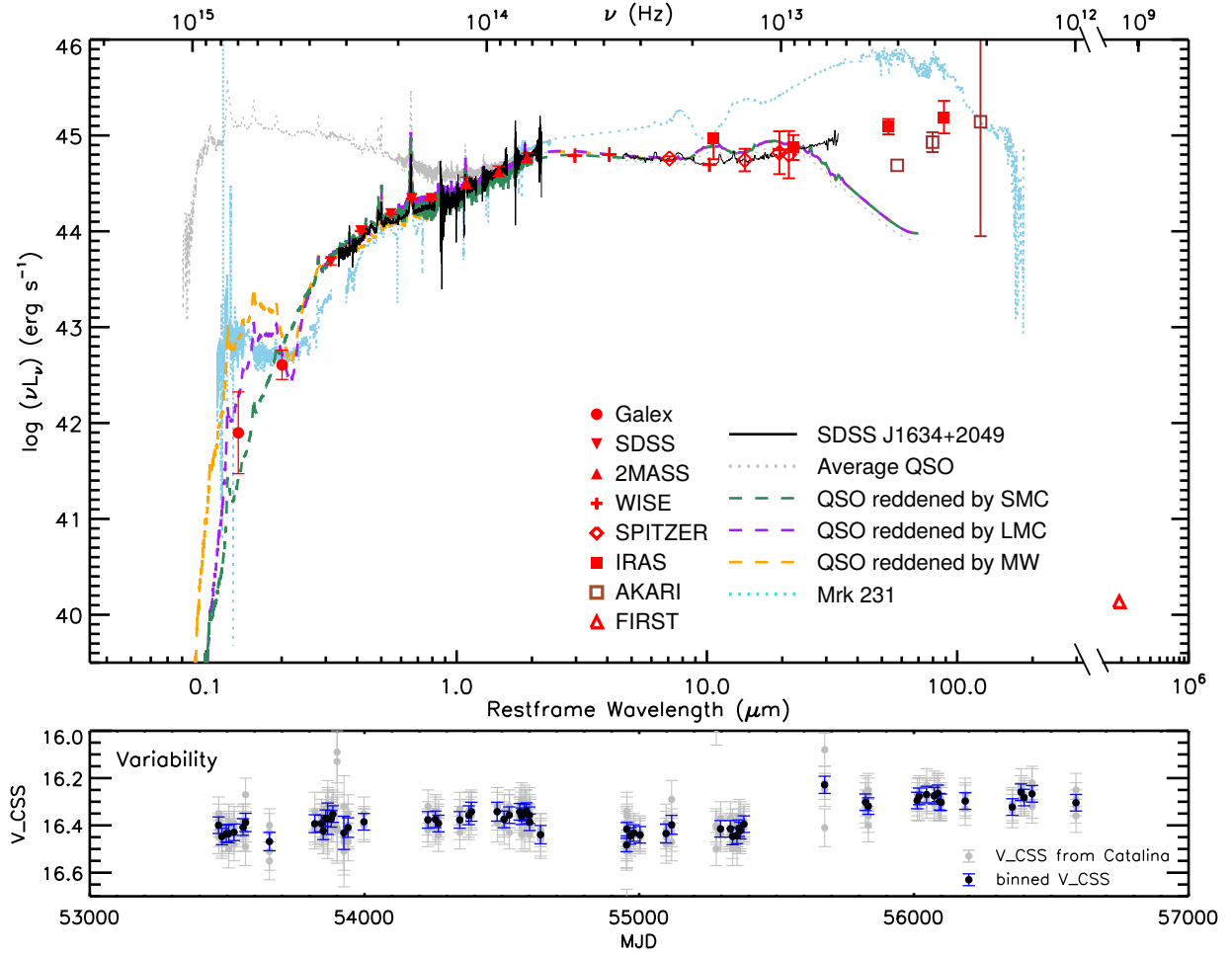}
      \caption{\footnotesize
    Top: The broadband spectral energy distribution of \thisobj\ constructed
    from non-simultaneous photometry data (red) and spectra, corrected for Galactic
    reddening and brought to the restframe.
    The black solid lines are the SDSS spectrum, the NIR spectrum observed by
    P200 TripleSpec, and the $Spitzer$ IRS spectrum.
    Overplotted for comparison is
    the average QSO spectrum (gray dotted line; see the text);
    the reddened versions of the average QSO spectrum
    by different extinction curves
    are denoted by the orange (MW), purple (LMC), and green (SMC)
    dashed lines.
    %reddened by Milky way \citep{1999PASP..111...63F}, LMC \citep{1999ApJ...515..128M}
    %and SMC \citep{1982ApJ...255...70H} extinction curve with $E(B-V)$ = 0.64,
    %0.66 and 0.61 respectively.
    The light blue doted line is the SED of Mrk~231 constructed from the multiband
    spectra scaled at $\sim 2 \micron$.
    Bottom: the variability of the $V$ band of \thisobj. The raw data (gray dots) are taken from
    Catalina Survey with the binned version within every day plotted also (black dot).
    \label{fig:sed}}
   \end{figure}

   As shown in Figure~\ref{fig:sed}, we construct the broadband SED in rest frame 
   wavelength using the photometric data and spectra in multiple bands.
   These data are corrected for Galactic extinction using the dust map of 
   \citet{1998ApJ....500..525S} and the \citet{1999PASP..111...63F} reddening curve.
   Because these photometric and spectroscopic observations are non-simultaneous,
   we first check the variability of this object before we analyze the broadband SED.
   The Catalina Sky Survey\footnote{The website is http://nesssi.cacr.caltech.edu/DataRelease/.
   The Catalina Sky Survey (CSS) is funded by the National 
   Aeronautics and Space Administration under grant no. NNG05GF22G issued through 
   the Science Mission Directorate Near-Earth Objects Observations Program.  
   The CRTS survey is supported by the U.S.~National Science Foundation under
   grant no. AST-0909182.}
   performs an extensive photometric monitor since 2005 April 9
   (MJD from 53469 to 56590), and has 272 observations so far.
   We obtain these data from the Catalina Surveys Data Release 2 (CSDR2), and
   bin it every day.
   As the bottom panel of Figure~\ref{fig:sed} shows, \thisobj\ has a long-term optical 
   variability within 0.2 mag in the $V$ band.
   Such a variability amplitude does not impact our discussions on its SED and so on below.
   %-2015June21 ---

   The top panel of Figure~\ref{fig:sed} shows the broadband SED of \thisobj.
   The average QSO spectrum from the UV to FIR band scaled at 2$\micron$
   is overplotted for comparison.
   This average QSO spectrum is combined from the UV to optical average 
   QSO spectrum of \citet{2001AJ....122..549V}, the NIR average QSO spectrum 
   of \citet{2006ApJ...640..579G} and the FIR average QSO spectrum of 
   \citet{2007ApJ...666..806N}. 
   It is obvious that the observed SED of \thisobj\ shows a very different shape 
   from that of the average QSO spectrum.
   In the UV, optical, and NIR $J$ and $H$ bands, \thisobj\ is much lower than the 
   average QSO spectrum;
   from the $K$ band up to the MIR (5--30~$\micron$), the shape of the SED is
   similar to that of the average QSO spectrum;
   in the FIR band, \thisobj\ shows an obvious excess, 10 times higher than
   the luminosity of the average SED of QSOs at $60\micron$.
   
   %% to create a new paragraph: --- 2016Feb24 ---
   Note that in Figure~1 the $AKARI$ photometric flux densities are systematically 
   lower than the $IRAS$ ones, which is actually because the $AKARI$ data at 65\micron\ and 140\micron\ 
   are not reliable.
  For the AKARI data, the quality flags at 65\micron, 90\micron\ and 140\micron\ are ``1'', ``3'' and ``1'', respectively,
  where ``3'' indicates the highest data quality and ``1'' indicates that the source is not confirmed.
   For the $IRAS$ data, the quality flags of 
  the flux densities at 12\micron, 25\micron, 60\micron, and 100\micron\ are ``1'', ``3'', ``3'' and ``2'', respectively,
  where  ``3'' means the highest data quality and ``1'' means that the flux is only an upper limit. 
    Thus, the $IRAS$ flux densities at 25\micron, 60\micron\ and the AKARI flux density at 90\micron\ are the most reliable; 
  the $IRAS$ flux density at 100\micron\ is the second most reliable with a quality flag of ``2'',
  and it is consistent with the $AKARI$ flux densities at 90\micron\  within 1-$\sigma$.
  The $IRAS$ 12\micron\ flux density is higher than the $WISE$ $W3$ (12\micron), which is because 
  the $IRAS$ 12\micron\ datum is just an upper limit  (with a $IRAS$ quality flag of ``1'').
  Besides, the flux density of the $IRAS$ 25\micron\ is consistent with those of the  
  $WISE$ $W4$ and $SPITZER/IRS$ 22\micron; 
  the flux densities of $WISE W3$ (12\micron) and $W4$ (22\micron) are consistent with those of $SPITZER/IRAC$ 8\micron, 
  $SPITZER/IRAC$ 16\micron, and $SPITZER/IRS$ 22\micron.

   %% to divide a new paragraph here: --- 2016Feb24 ---
   We try to match its SED to the reddened versions of the average QSO spectrum with different
   extinction curves.
   In Figure~\ref{fig:sed}, the blue dashed line indicates the average QSO spectrum 
   reddened with Milky Way extinction curve \citep{1999PASP..111...63F} by $E_{\rm B-V}$ = 0.64,
   while the purple and green dashed lines show the reddening with SMC extinction
   curve \citep{1992ApJ...395..130P} by E$_{\rm B-V} = 0.61$ and the LMC extinction curve
   \citep{1999ApJ...515..128M} by $E_{\rm B-V}$ = 0.66, respectively.
   Here, the $R_{\rm V}$ for the Milky Way extinction curve is 3.1, and for the LMC 
   extinction curve is 2.6 \citep{2001ApJ...548..296W}.
   It is hard to distinguish the extinction types according to the optical and
   NIR spectra, since these reddened average QSO spectra show little difference
   in the optical and NIR bands.
   Considering also the NUV and FUV photometric data retrieved from the GALEX 
   archive (albeit not as superior as a UV spectrum),
   the LMC extinction curve is favored (see Figure~\ref{fig:sed}).
   Hereafter, we will employ the LMC extinction curve with $R_{\rm V} =2.6$ for the 
   internal dust obscuration of  \thisobj.
    %-2015June21 ---

   %% IR and SFR ---
   The IR luminosity (8-1000 $\micron$) is calculated based on the
   $IRAS$ photometric fluxes following \citet{1996ARA&A..34..749S}.
   %% 2016Feb24 ---:
   Because  the $IRAS$ 12\micron\ flux density  is an upper limit,
   in the calculation  
   we use the $WISE$ datum in the $W3$ band (12\micron) instead. 
   It gives log$L_{\rm IR}(\lsun) = 11.91\pm0.03$, which is very closed to the
   defining IR luminosity of ULIRGs.
   As \citet{2006ApJ...649...79S} suggested, for typical QSOs (e.g., PG~QSOs),
   most of the far-infrared luminosity is originated from star formation.
   \rev{If this is the case for \thisobj, following the relation 
     SFR$(\msun yr^{-1}) = 4.5\times 10^{-44} L_{\rm IR}$ (\ergs)
   \citep{1998ARA&A..36..189K}, the star formation rate (SFR) is estimated to be 
   SFR $= 140 \pm 43 \msun$~yr$^{-1}$.
   The scatter of this equation is $\pm$ 30\% \citep{1998ARA&A..36..189K}, which is dominated 
   the statistical uncertainty of the SFR.}
   From the polycyclic aromatic hydrocarbon (PAH) emission lines and 24$\micron$ continuum, 
   the SFR are estimated to be $141 \msun$~yr$^{-1}$ and $143 \msun$~yr$^{-1}$ respectively, 
   which are well consistent with the SFR estimated from its IR luminosity.
  
   %% start a new paragraph: radio ----
   The $k-$corrected radio power at 1.4GHz for \thisobj\ is also estimated, %%to be
   $P_{1.4\rm GHz} = 9.14 \times 10^{23}$ W Hz$^{-1}$.
   This is calculated by $P_{1.4\rm GHz} = 4\pi D_{\rm L}^2 f_{\rm int}/(1+z)^{1+\alpha_{\rm r}}$,
   where the radio spectral index $\alpha_{\rm r}~(F_{\nu} \varpropto \nu^{\alpha_{\rm r}})$ is assumed 
   to be $-$0.5. 
   
   %% start a new paragraph: Mrk231 ----   
   We compare the SED of \thisobj\ with that of Mrk~231, which is a prototypal 
   nearest ULIRG/quasar composite object.
   The SED of Mrk~231 is constructed from the multiband spectra we collected.
   The FUV (1150--1450\AA) spectrum is obtained with $Cosmic\  Origins\  Spectrograph$ (COS)
   G130M grating on board the $Hubble\ Space\  Telescope$ (HST).
   The UV spectrum within (1600--3200\AA) is obtained with $Faint\  Object\  Spectrograph$ (FOS)
   G190 and G270 gratings on board HST.
   All the HST spectra are retrieved from HST data archive.%%-
   \footnote{http://archive.stsci.edu/hst/search.php}
   The optical spectrum (3750--7950 \AA) is obtained from \citet{1995ApJS...98..129K}.
   We also observed Mrk~231 using TripleSpec spectrograph on Hale telescope on 2013 February 23
   to obtain its NIR spectra.
   The FIR spectrum is obtained from \citet{2008ApJS..178..280B}.
   These multibands spectra are corrected for the Galactic extinction,
   and combined together by matching to its photometry data.
   %\footnote{This composite spectrum of Mrk~231 is available on request to Wen-Juan Liu (zoey@mail.ustc.edu.cn).}
   In Figure~\ref{fig:sed} we scale the SED of Mrk~231 at 2$\micron$ to that of \thisobj.
   As Figure~\ref{fig:sed} demonstrates, the shapes of the two SEDs differ significantly.
   Mrk~231 appears more reddened in the NUV--optical continuum.
   More strikingly, Mrk~231 has a much larger excess in MIR and FIR bands 
   with a deeper silicate absorption dip at 9.7$\mu$m than \thisobj.
   
   %-2015June22 ---

    \subsection{Analysis of SDSS Images \label{subsec:sdssimage}}

\begin{figure}[htbp] 
      \center
      \includegraphics[width=5in]{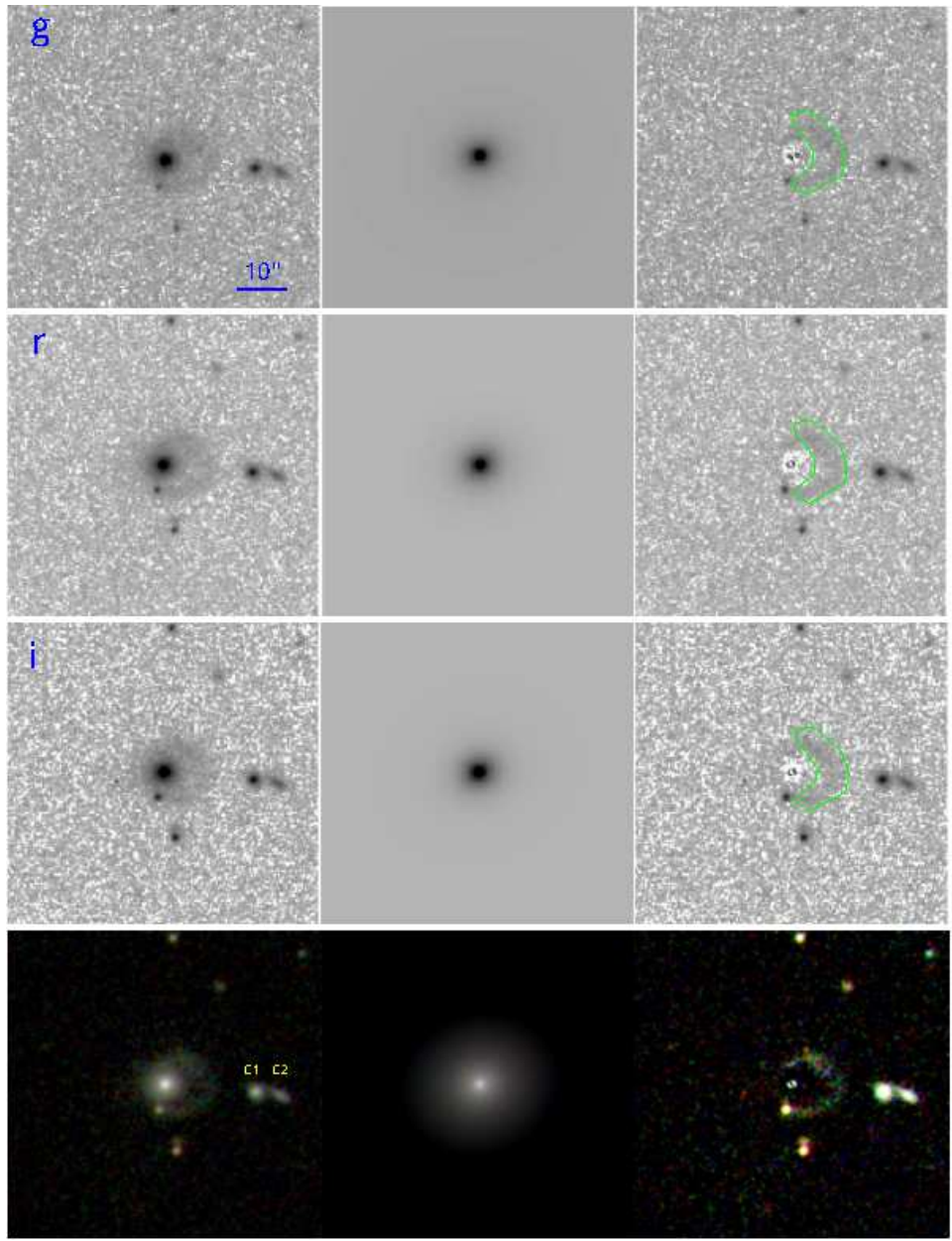}
      \caption{SDSS images of \thisobj\ and the 2D imaging decomposition
      by GALFIT. From top to bottom are SDSS $g$, $r$, and $i$ bands, and the composite
      of the three bands, respectively. The left column shows the original
      image, the middle column shows the GALFIT model (PSF + S\'{e}rsic), and the
      right column shows the residual image. All images are oriented with north up
      and east to the left; the black line marks a scale of 10\arcsec ($\sim$ 23.1 kpc).
      The green polygon denotes the ring region, which as well as the foreground stars 
      has been masked out in GALFIT fitting. \label{fig:2dimage}}
   \end{figure}

     \thisobj\ was photometrically observed by SDSS in the $u, g, r, i$, and $z$ bands on
     2003 June 23 UT, with an exposure time of 54~s per filter in drift-scan mode
     \citep{1998AJ....116.3040G}.
     A bright point-like source appears at the center of its images and the whole galaxy
     is almost round in shape and has no spiral arms, indicating an elliptical/spheroidal
     or a face-on S0 galaxy.
     Closer inspection reveals a low surface brightness (SB) yet visible, circumgalactic
     ring-like structure.
     Although the standard SDSS images have a relatively short exposure time and low spatial
     resolution due to the seeing limit, they are still very useful for global study.
     % in the way of 
     % analyzing the 1-dimensional, azimuthally averaged SB profile (cf. Jiang~\etal\ 2013).
     %%-- the above omitted. -xbdong, 2015June23
     The drift-scan mode, yielding accurate flat-fielding, in combination with the large
     field of view (FOV) ensures very good measurement of the sky background, and thus the
     azimuthally averaged, radial SB profile can be reliably determined down to
     $\mu_r \approx27$ mag~arcsec$^{-2}$
     %(e.g., Pohlen \& Trujillo 2006, Erwin~\etal\ 2008, Jiang~\etal\ 2013).
     \citep[e.g.,][]{2006A&A...454..759P,2008AJ....135...20E,2013ApJ...770....3J}.
     We try to perform a two-dimensional (2D) decomposition of the AGN and host galaxy
     of \thisobj\ using GALFIT \citep{2002AJ....124..266P,2010AJ....139.2097P}.
     An accurate decomposition cannot only help us understand its host galaxy, but also
     put an independent constraint on the SED of the AGN, which is helpful for spectral
     fitting \citep[cf.][]{2013ApJ...770....3J}.
     Taking advantage of the large FOV, a bright yet unsaturated nearby star
     (SDSSJ163503.44$+$204700.3) is selected as the PSF image, whose precision is 
     essentially important to separate the AGN from the host.
     The host galaxy is represented by a \citet{1968adga.book.....S} $r^{1/n}$ function.
     During the fitting, the sky background is set to be free; the outer ring region,
     enclosed by the green polygon as in Figure~\ref{fig:2dimage}, is masked out.
     All other photometric objects in the field identified either by
     Sextractor \citep{1996A&AS..117..393B} or by the SDSS photometric pipeline are
     also masked out.

\begin{figure}[htbp]
      \center
     \includegraphics[width=4in]{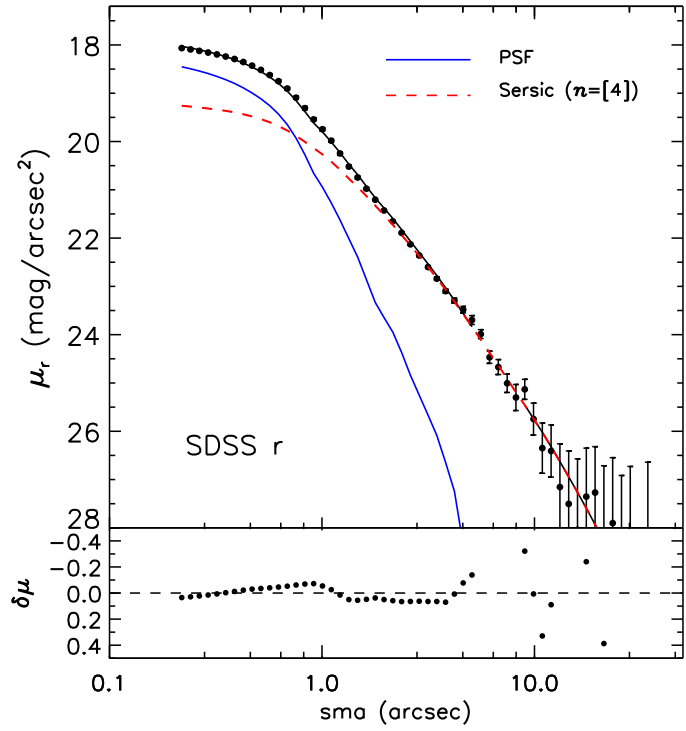}
      \caption{Top: one-dimensional representation of the two-component GALFIT model
    applied to the SDSS $r$ image of \thisobj: PSF for the nucleus
    (blue solid line) $n=4$ S\'{e}rsic function for the host galaxy (red dashed line).
    The sum of the two components is shown as the black solid line. The observed data
    are plotted as black dots with $\pm 1\sigma$ error bars.
    Bottom: the residuals between the observed data and the best-fit model.
    \label{fig:1dmodel}}
   \end{figure}

     We begin the fitting with a free PSF~$+$~\sersic\ scheme allowing all
     parameters to vary, which yields an unreasonable high value of \sersic\ index ($n>10$).
     Then we thus try to fixed $n$ to 4, 3, 2, and 1, respectively \citep[see][]{2013ApJ...770....3J}.
     Except for the $u$-band image, which is totally dominated by the AGN component, all other
     four bands yield well consistent results: the best-fit \sersic\ component is with $n=4$.
     We have also attempted to add an exponential disk component, yet no convergent result 
     can be achieved. 
     The fitting results in the $g$, $r$, and $i$ bands are summarized in Figure~\ref{fig:2dimage} 
     and Table~\ref{tab:galfit}.
     To illustrate the AGN and host galaxy contribution at different radii, shown in
     Figure~\ref{fig:1dmodel} shows
     the corresponding radial SB profiles of the best-fit components of the $r$ band image.

     As we know, the SDSS spectrum is extracted through a fiber aperture of 3\arcsec\ in diameter.
     To further assess the AGN and host galaxy fluxes in the 
     fiber aperture, we have also integrated the fluxes within the aperture from the model images
     of the AGN and \sersic\ components, respectively, which are also listed in Table~\ref{tab:galfit}.
     %% --2015June23 ---
 
\subsection{Analysis of the Optical-NIR Spectrum}

   \subsubsection{Decomposition of the continuum \label{subsec:continuum}}

   Before we perform the analysis on its spectra from different telescopes/instruments, 
   we first check the aperture effect because \thisobj\ is an extended galaxy.
   Although the SDSS, DBSP, and NIR spectra are observed with different apertures/slits, 
   we find that the three spectra, which were calibrated independently, are well consistent 
   in flux level; especially, DBSP, and SDSS spectra are almost the same 
   in their overlap part.
   On the other hand, as Figure \ref{fig:1dmodel} demonstrates, the surface brightness of 
   \thisobj\ decreases rapidly in $r >$1\arcsec, and therefore the outer region ($r >$1\arcsec.5)
   has a negligible contribution to these spectra.
   %as the imaging decomposition shows (see Figure \ref{fig:1dmodel}), 
   %the host galaxy is of low SB compared to the AGN emission.
   %Therefore the outer region ($r >$1\arcsec.5) has a negligible contribution 
   %to these spectra.
   %the aperture effect caused by the host-galaxy starlight ($r >$1\arcsec.5)
   %is not significant.
   %%-- revised a bit the following sentence: --2016Feb24 --
   Besides, considering the above fact that the SDSS and DBSP spectra are almost the same
   although the three spectra were observed in different time (SDSS: MJD 53224,
   NIR: MJD 56034, DBSP: MJD 56771),
   the variability effect can be ignored.
%%2015June26 night---

\begin{figure}[tbp]
    \center
    \includegraphics[width=5.2in]{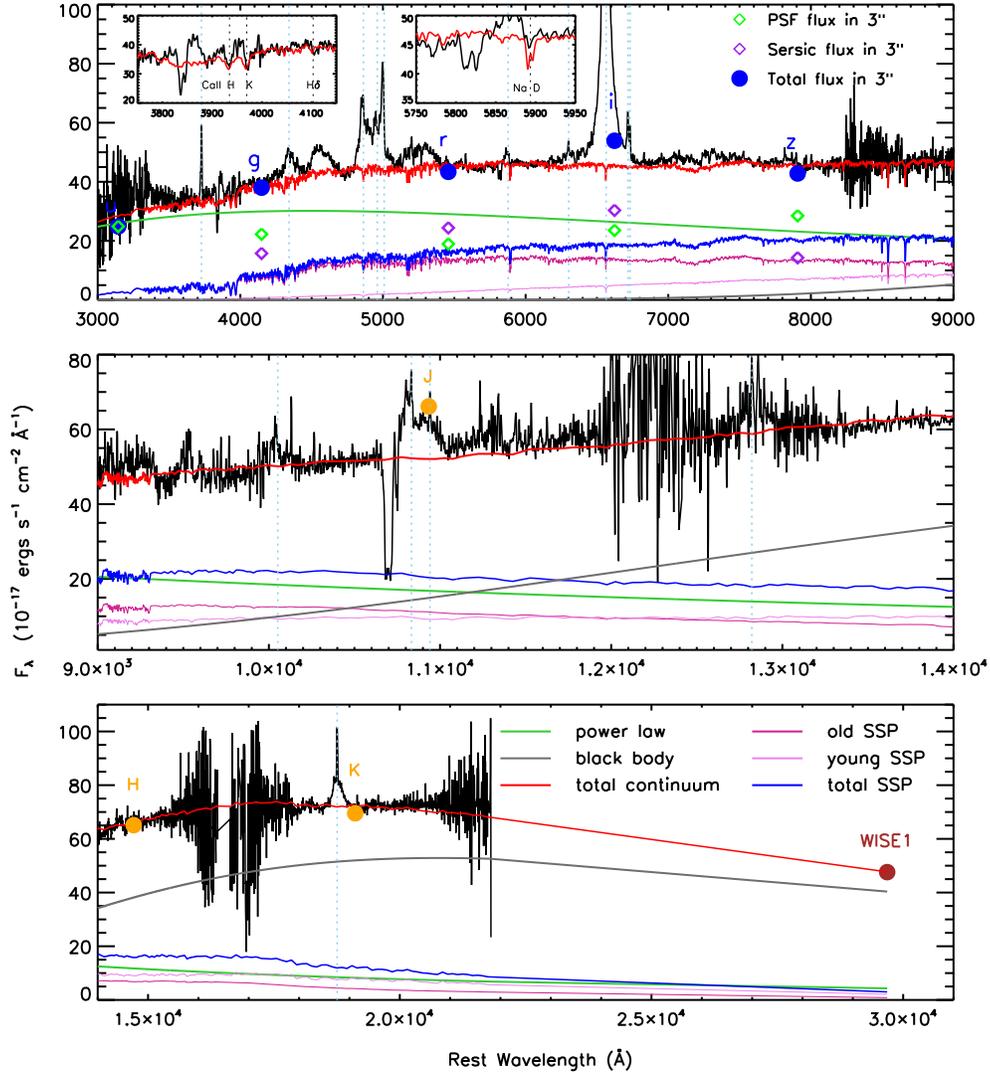}
    \caption{\footnotesize Detailed decomposition for continuum of \thisobj\
      in the rest frame wavelength range from 3000~\AA\ to 3~$\micron$.
      The black solid line is the combined spectrum of SDSS (3350-8150~\AA), 
      DBSP (2900-9200~\AA) and TripleSpec spectrum (8600~\AA-2.2~$\micron$). 
      The common part of the three spectra were weighted by spectral signal to noise ratio(S/N).
      The blue solid dots are fluxes of $u$, $g$, $r$, $i$, and $z$ bands in 3$\arcsec$ measured from
      SDSS images (see \S\ref{subsec:sdssimage}).
      The green and purple diamonds shows the decomposition of PSF and S\'{e}rsic components in 
      3 $\arcsec$ decomposed from SDSS images respectively.
      Orange and brown dots are the photometry fluxes from 2MASS $J$, $H$, $K$, and $WISE$ $W1$ bands.
      All the spectrum and photometry data are corrected for Galactic reddening and redshift.
      Green solid line represents the power-law component reddened by LMC extinction curve with
      $E_{\rm B-V}$ = 0.41, and the gray line represents the black body continuum from hot dust 
      (torus) with a temperature of 1394 K.
      The magenta line shows the continuum of the old stellar population (age $\sim$ 9 Gyr), 
      the violet line shows the heavily reddened continuum of the young stellar population 
      (age $\sim$ 127 Myr; $E_{\rm B-V}$ = 2.2), and the blue solid line shows the sum flux of 
      the old and the young stellar population. \label{fig:conti}}
 \end{figure}

   Figure~\ref{fig:conti} shows the spectrum combined from the DBSP, SDSS, and
   NIR spectra, the overlap parts of which are weighted by spectral S/N.
   First, we take a global overview of the features of the AGN and 
   starlight components.
   %analyze the composition of the observed spectrum for this object.
   In \S \ref{subsec:sdssimage} the 2D imaging decomposition by GALFIT gives 
   the contributions of the two components.
   Within the 3\arcsec-diameter aperture, the PSF (AGN) component accounts for 
   59\% of the $g$-band light, 44\% in the $r$, 44\% in the $i$, and 67\% in the $z$, 
   respectively.
   The u-band image is much less sensitive, so an accurate decomposition
   is difficult; our rough decomposition shows that it is totally dominated
   by a PSF component.
   According to the 2D decomposition, the colors ($g-r$, $r-i$) of the \sersic\
   component are (0.93, 0.5), which suggest that the age of the dominating
   stellar component could be older than 10~Gyr \citep{2003MNRAS.344.1000B}.
   Now looking at the combined spectrum,
   the typical AGN features such as strong broad emission lines and
   blueshifted absorption lines are significant;
   in contrast, the starlight component is almost lost of any features, except for the
   appearance of weak \caii~H\,\&\,K and \nai~D absorption lines.
   On the other hand, the high FIR luminosity betrays recent violent star formation 
   activities (see \S2.2), suggestive of the presence of a (obscured) young stellar population.
   Meanwhile, both the large extinction in the UV and the large Balmer decrement 
   of narrow hydrogen emission lines (see \S\ref{subsec:emissionline} and Table~\ref{tab:decrement}) 
   suggest the star formation region is enshrouded by thick dust.
   %% below revised on 2016Feb24 : ---
   Here we can roughly estimate the lower limit to the extinction for the young stellar 
   population that ionizes the H~II region and powers the nebular emission lines and FIR dust emission.
   Using the SFR estimated from PAH (see \S\ref{subsec:mirspec}) and the 
   relation $SFR(\msun \rm yr^{-1}) = 7.9 \times 10^{-42} L(\rm H\alpha)$ (\lum) 
   \citep{1998ARA&A..36..189K}, we get the predicted luminosity of \ha\ for the HII region,
    $1.78 \times 10^{43}$ \ergs.
   Yet the observed luminosity of the narrow \ha\ line $L$ (narrow H$\alpha$) is only 
   $1.88\times 10^{41}$ \ergs.
   Assuming all the emission of the narrow  \ha\ component is from star formation,
   then the extinction for narrow \ha\ emission line is $A_{\ha}$ = 4.9;
   applying the LMC extinction curve, the $E_{\rm B-V}$ of the young stellar population is thus 
   $\sim$2.6.
   In addition, the excess emission between $\sim2$ and 10~$\micron$, which
   appears in this spectrum, has been widely regarded to be originated from the hot
   dust of $\sim 1500$~K in the studies of AGN SEDs
   \citep{1978ApJ...226..550R,1986ApJ...308...59E,1987ApJ...320..537B,1994ApJS...95....1E,2006ApJ...640..579G}.

   Based on the above analysis, we can decompose the continuum (3000--22000~\AA\ in 
     rest frame wavelength) with the following model:
   \begin{footnotesize}
      \begin{equation}
	\begin{split}	
		F_{\lambda} = C_{\rm nucleus} \, A_{\rm nucleus}(E^{\rm nucleus}_{\rm B-V}, \lambda) \,  \lambda^{\alpha} ~ + ~ 
                          C_{\rm bb} B_{\lambda}(T_{\rm dust}) ~ + ~  \\
                          C_{\rm host,1} \, SSP(\geqslant 2 {\rm Gyr}) ~ + ~ 
			   C_{\rm host,2} \, A_{\rm host}(E^{\rm host}_{\rm B-V}, \lambda) \, SSP(\leqslant 1 {\rm Gyr}) ~,
         %% -- 2016Feb24 ---
	  % F_{\lambda} =  C_{\rm nucleus}A_{\rm nucleus}(E^{\rm nucleus}_{\rm B-V}, \lambda)\,\lambda^{\alpha} + 
	  %              C_{\rm bb}B_{\lambda}(T_{\rm dust}) + C_{\rm host,1}SSP(\geqslant 2 {\rm Gyr})+ \\
	  %		 C_{\rm host,2}A_{\rm host}(E^{\rm host}_{\rm B-V}, \lambda) SSP(\leqslant 1 {\rm Gyr}) ~,
          %  % C_{\rm bb}B_{\lambda}(T_{\rm dust})  ~,
	 \end{split}
      \end{equation}
    \end{footnotesize}
   where $F_{\lambda}$ is the observed spectrum in the rest frame,
      %%%2016Feb24 below: -------
   $C_{\rm nucleus}$, $C_{\rm host,1}$, $C_{\rm host,2}$, and $C_{\rm bb}$ are 
   the factors for the respective components; $A_{\rm nucleus}$ and $A_{\rm host}$ are the 
   dust extinction to the AGN emission and to the young stellar population, respectively;
   and $B_{\lambda}(T_{\rm dust})$ is the Planck function.
   %%% slope and E(B-V), 2016Feb24 ----
   As $\alpha$ (the intrinsic AGN continuum slope) and  $E^{\rm nucleus}_{\rm B-V}$
   are somehow degenerate, in the fitting $\alpha$ is fixed to $-1.7$
   while $E^{\rm nucleus}_{\rm B-V}$ is free,
   which is the common recipe for reddened AGN continua in the literature
   (see, e.g., Dong~\etal\ 2005, 2012, and Zhou \etal\ 2006).
   The two stellar populations are modeled by two simple stellar
   population (SSP) templates from \citet{2003MNRAS.344.1000B},
   with the metallicity being fixed to the solar ($Z=0.02$) for simplicity.
   %% -- the definition of SSP : 2016Feb24 ---
   \rev{A SSP is defined as a stellar population 
   whose star formation duration is short compared with the lifetime of its most massive stars.}
   Based on the analysis in the preceding paragraph, we select 30 SSP
   templates with ages between 50~Myr and 1~Gyr to model the young stellar
   component, and 28 SSP templates with ages between 5~Gyr and 12~Gyr to model
   the old stellar component. 
   We traverse every possible SSP template in the library during the fitting, 
   %%2016Feb24
   and use the IDL routine MPFIT \citep{2009ASPC..411..251M} to do the job with 
   $C_{\rm nucleus}$, $C_{\rm host,1}$, $C_{\rm host,2}$, $C_{\rm bb}$,
   $E^{\rm nucleus}_{B-V}$, $E^{\rm host}_{B-V}$ and $T_{\rm dust}$ being
   free parameters.
   Detailedly, as the average QSO spectrum reddened with LMC extinction curve
   by $E_{B-V} =0.66$ fits this object well
   (see Figure~\ref{fig:sed}), the $E^{\rm nucleus}_{B-V}$ is initially set
   to be 0.66 and allowed to vary freely within 0.3--0.8 according to the
   Balmer decrement measured from the broad-line \ha/\hb;
   The $E^{\rm host}_{B-V}$ is initially set to be 2.6, varying within 1.0--3.0
   according to the Balmer decrement measured from the narrow-line \ha/\hb\
   (see Table~\ref{tab:decrement}).
   We also set the fraction of the power-law component in the $r$ band to be 0.5 initially
   and allow the value varying within 0.4--0.6.
   %Beside the spectrum, we also add the photometric data WISE W1 (3.3$\micron$) when fitting, as 
   %it is helpful to constrain the red end of the blackbody.

   The fitting converges on the two SSP templates with ages of 127~Myr and 9~Gyr,
   respectively, $E^{\rm nucleus}_{B-V}$ of 0.41, $E^{\rm host}_{B-V}$ of 2.2, 
   and $T_{\rm dust} =$1394~K that is close to the result of \citet{2006ApJ...640..579G}.
   As Figure~\ref{fig:conti} shows, the fractions of the decomposed nucleus and starlight 
   components in the spectrum are basically consistent with the imaging decomposition. 
   The blackbody continuum from the hot dust of the presumable torus should be 
   the dominant emission in the $K$ band and the $WISE$ $W1$ band.
   To check it, we extend the fitting result to the $WISE$ $W1$ band, and find that it can well
   reproduce the observed data point.
   As to the best-fit starlight spectrum composed of the two SSP components,
   it reproduces well the \caii~H\,\&\,K and \hd\ absorption lines, yet it overestimates the
   \nai~D absorption.
   This discrepancy may be due to the uncertainty of the decomposition or more likely
   the contamination to the \nai~D absorption by the nearby \hei$\lambda5876$ emission line.

   %%-start a new paragraph---
   
   The concern on the decomposition is mainly about the decomposition of the two
   stellar components.
   %% comment on the use of TWO ssps: --- 2016Feb25 ---
   To assess the reliability of the decomposed stellar components (mainly of their ages),
   we do the following checks.
   First, regardless of the above physical arguments to justify the use of {\em two} SSPs,
   we test this point with the data (lest the spectral quality should not be sufficient to support it).
   Using a single SSP with the metallicity and age being free parameters for the starlight in the model,
   we obtain the best fit with a much larger minimum \chisq.
   The reduced \chisq\ increases by
    $\Delta \chisq_{\nu} = 2.5$, and the two-SSP model is favored according to $F-$test
   (the chance probability $\pnull = 0$). 
   Then we check the algorithm of MPFIT for global minimum, as follows.
   (1) We loop over the grid of the 28 templates for the old stellar population component
   and for every case of the assigned template for the old component
   we get the best-fit young stellar component according to minimum \chisq;
   these 28 best-fit young populations have ages in the range of 80--210~Myr.
   (2) On the other hand, we loop over the grid of the 30 templates for the young stellar population component
   and for every case we get the best-fit old component by \chisq\ minimization;
   these 30 best-fit old stellar populations have ages in the range of 8--11~Gyr.
   We can see that, at least, the ages of the two components can be well separated.
   (3) Furthermore,
   considering that the \caii~H\,\&\,K and \hd\
   absorption lines are dominated by the old stellar component,
   we devise instead a \chisq$_{\rm abs}$ as calculated in the spectral region of 3900-4050\AA\
   in order to better constrain the fitting of the old stellar component.
   We repeat the procedure of (2) yet with minimizing \chisq$_{\rm abs}$.
   For every case of the held template for the young component,
   the best-fit SSP template for the old component may be different from (2).
   Yet the ages of the best-fit old components of all the 30 cases are still in the range of 8--11~Gyr.
   %For each group models that adopt the same young stellar SSP template,
   %we select the model with the minimum \chisq\ of the whole fitting and the model with
   %the minimum \chisq$_{\rm abs}$.
   %Though the ages of old stellar component corresponding to above two situations are not
   %exactly the same, all of them are between 8 and 11 Gyr.
   %However, the decomposition still have large uncertainty due to the uncertainty of the
   %dust extinction of different components.
   Certainly, we should note that the above checks do not account for the effect of the 
   internal dust extinction parameters (see Eq.~1), which should impact the fitting of the 
   two stellar components.

   %% add a new paragraph, 2016Feb25: ---- 
   Since for any a SSP the age and metallicity are degenerate parameters, 
   we also try to model the starlight 
   with two SSPs with their age and metallicity set to be free
   (allowing $Z$ = 0.0001, 0.0004, 0.004, 0.008, 0.02 and 0.05).
   This test scheme yields that the best-fit two SSPs have ages of 47.5~Myr and 12~Gyr, 
   and both converge to an extremely low metallicity $Z = 0.0004$.
   Such a metallicity is even lower than that of most of the most metal-poor (dwarf) galaxies,
   which is unrealistic for \thisobj\ with a mass/size similar to the Milky Way.
   %%LWJ: As we know, high mass early type galaxies tend to have much higher metallicity than this. 
   %% We also find the best-fit of this model has only slightly better normalized \chisq\ than 
   %% the one with the metalicity of the SSPs fixed to be the solar. 
   Besides,
   this scheme does not change much the best-fit starlight (the sum of the two stellar populations):
   the difference of the starlight between this scheme and the solar metallicity scheme we adopt
   is 14\% in the \hb---\oiii\ region, and only 2\% in the $K$ band;
   the best-fits of the other components (the reddened AGN continuum and the hot dust emission)
   change negligibly also.
   %% the below is omitted---- 
   % LWJ: therefore different metallicity does not affect the estimation for stellar mass using mass-to-light 
   % LWJ: ratio $M/L_{\rm K}$ (see \ref{subsec:ring}. 
   % Taken all above, the spectral quality of \thisobj\ at hand is not high enough
   %for us to accurately determine the metallicity and age for host galaxy. 
   In this present work, we use $K$-band luminosities to derive stellar masses (see Sections~3.1 and 3.2) 
   and do not use stellar age and metallicity to achieve any conclusions.
   Thus, it is safe to assume the solar metallicity for the two SSPs in the model.

 %-- S2.4.2 :-----
 \subsubsection{Emission lines \label{subsec:emissionline}}

 As the Figure~\ref{fig:conti} shows, strong emission lines of \thisobj\
  display intensively in four spectral regions: \hb\ + \oiii$\lambda\lambda4959,5007$ + \feii\ 
  multiplets, \ha\ + \nii$\lambda\lambda6548,6583$ + \sii$\lambda\lambda6716,6731$,      
 \pc\ + \hei$\lambda10830$ and \pa\ (note that the \pb\ and \pd\ emission
 lines are of low S/N).

\begin{figure}[htbp]
 \includegraphics[width=6.4in]{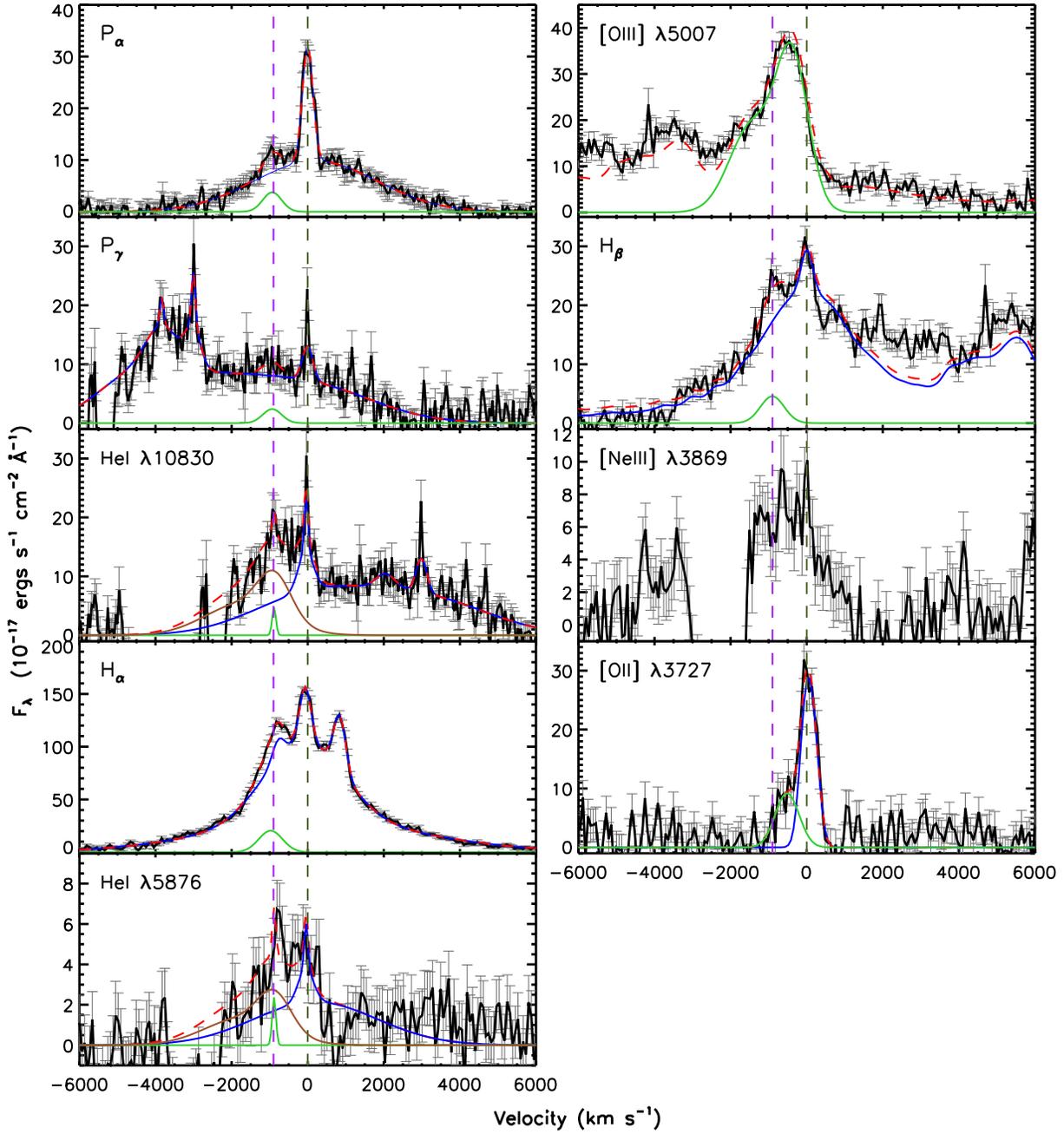}
   \caption{Demonstration of the profile of the various emission lines,
   particularly their blueshifted components (presumably caused by AGN outflow).
   Plotted are the observed spectra with continuum and \feii\ emission subtracted (black lines),
   the total fitted profile of each emission line (red dashed lines), 
   the sum %LWJ - whole profile 
   of the unblueshifted components of each line (blue lines), 
  the narrow blueshifted component (green lines),   
  and broad blueshifted component (brown lines). The purple and olive dashed vertical lines indicate 
	    $-900$~\kms\ and 0 \kms, respectively. \label{fig:outflow}}
\end{figure}

 Before we fit the four regions, we first take a look at the profiles of %different 
 the various emission lines.
 Figure \ref{fig:outflow} shows \pa, \pc, \hei$\lambda10830$, \ha, \hei$\lambda5876$, \oiii$\lambda5007$,
 \hb, \neiii$\lambda3869$ and \oiii$\lambda3727$ emission lines in the common velocity space.
 We note the following major points.  %%2016Feb26 ---
 (1) The total profiles of the recombination lines (such as \pa, \pc, \hei$\lambda10830$, \ha, \hei$\lambda5876$ and \hb)
 apparently exhibit a narrow peaked component and a lower and broad base;
 the two components appears separable from each other. 
 (2) Besides, probably of the most interesting, 
 in \hb\ and \hei$\lambda\lambda5876,10830$, there clearly exists
 an extra cuspy narrow component that is blueshifted by a velocity of  $\sim900$~\kms;
 the blueshift velocity is consistent in the three lines.
 %%% LWJ: an consistent obvious extra cuspy component blueshifted by a velocity of $\sim 900$ \kms.  
 %%
 (3) The whole profile of high-ionization narrow forbidden emission lines, 
 \oiii$\lambda5007$ and \neiii$\lambda3869$, is blueshifted evidently,
 with an offset velocity of $\approx$500~\kms\ according to the peak of \oiii$\lambda5007$;
 this kind of bulk blueshifting seems to be present in the low-ionization 
 %ed component appears  The low-ionization 
 forbidden line \oii$\lambda3727$,
 yet with a smaller blueshift. 
 %LWJ: also shows an obvious bulk-blueshifted component, which has a smaller blueshift
 %than in the aforementioned high-ionization lines.

%---------------------------------------------------------- Pa
 We start the line profile fittings with the \pa\ emission line,
 which is the strongest of the hydrogen Paschen
 series, and basically free of blending (unlike \hei\,$\lambda10830$).
 Its profile can be used as a template to fit the \pc\ $+$ \hei$\lambda10830$
 blends \citep{2008ApJS..174..282L}.
 The high contrast between the narrow peak and broad base of the \pa\ profile  
 makes it be able to decompose easily into the broad and narrow components. 
 There is another weak yet significant excess cusp on the blue side of the
 \pa\ profile. If this cusp belongs to \pa, its velocity offset
 is almost the same as the narrow cusp of \hb\ mentioned above 
 (at velocity of $-900$~\kms\ in Figure~5).
 Therefore, we take this cusp as the third component of the \pa\ profile in the model.
 Every components are modeled with (multiple) Gaussians.
 Initially, the narrow, the blueshifted cuspy, and the broad components are fitted 
 with one Gaussian each;
 more additional Gaussian(s) can be added into the model for a component 
 if the $\chisq$ decreases significantly with an $F$-test probability $\leq$0.05.
 A first-order polynomial is adopted to fine-tune the local continuum.

\begin{figure}[tbp]  
   \center
   \includegraphics[width=6.4in]{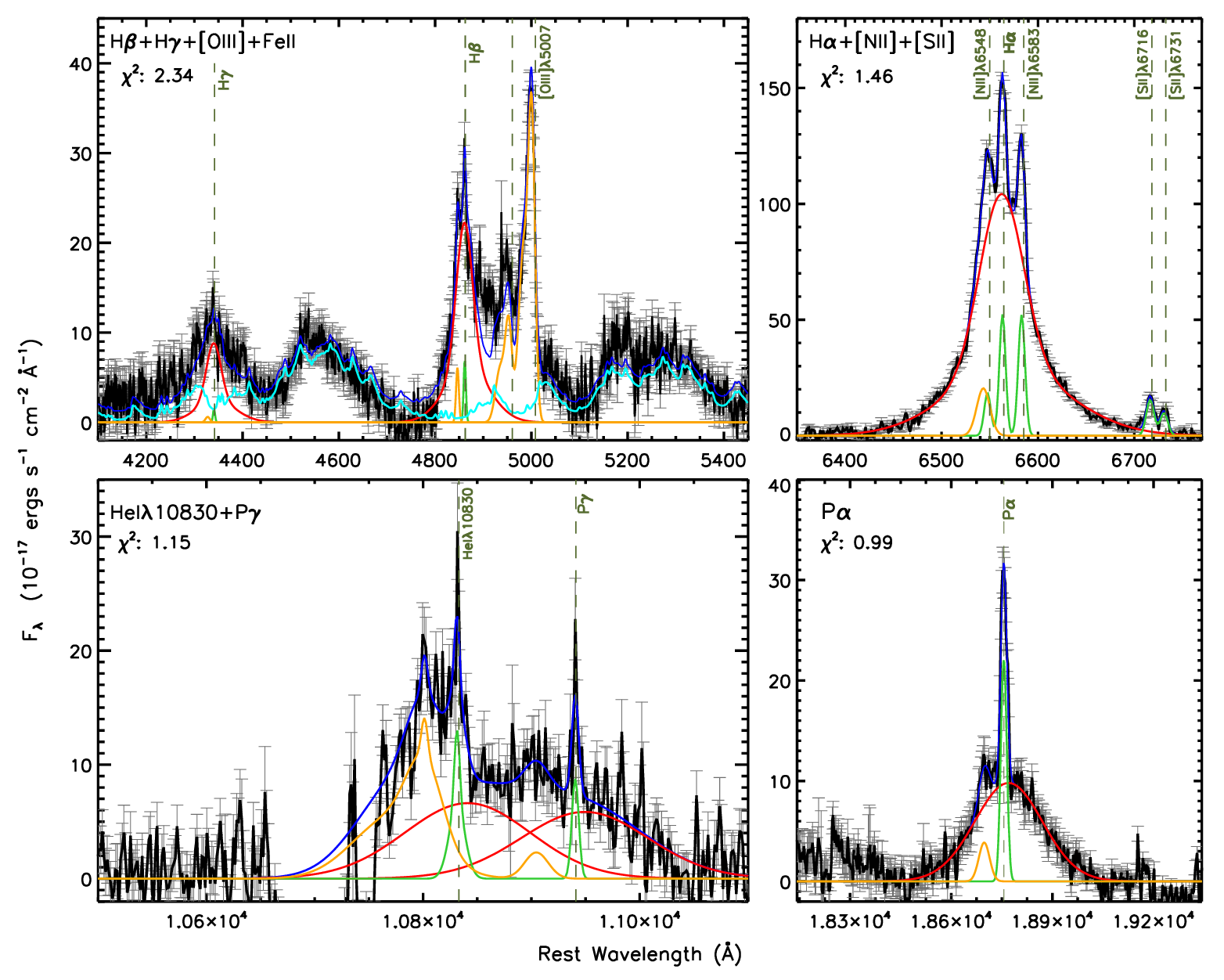}
   \caption{The best-fit models for \hb + \hc + \oiii + FeII region (upper left),
            \ha + \nii + \sii\ region (upper right), \hei$\lambda10830$+\pc\ region
        (lower left) and \pa\ region (lower right).
        The blue solid line represents the total fit profile for each region.
        The red and green lines show the decomposed profiles of broad and narrow
        emission lines respectively.
        The orange line represents the blueshifted lines, and the cyan line represents
        optical Fe\,II multiplets. \label{fig:linefit}}
  \end{figure}

 The fitting turns out,  with a reduced \chisq\ of 0.99, that  %%2016Feb26
 the narrow component and the blueshited cusp are well fitted with a single Gaussian each,
 and the broad component is sufficiently fitted with two Gaussians
 (see Figure~\ref{fig:linefit}).

 %------------------------------------------------------------ Pc + HeI 10830 ----
 The \pc\ emission line is heavily blended with the \hei\,$\lambda10830$ emission line. 
 In addition, two broad \heiabs$\,\lambda10830$ absorption troughs are located in 
 the blue wing of the \hei$\lambda10830$ emission line, which increases the 
 complexity of decomposing this blend.
 We use the \pa\ as the template to fit \pc.
 To be specific, we assume that 
%%2016Feb26: ----
 \pc\ has also three components, namely the narrow, broad, and blueshifted cuspy ones,
 and every component shares the same profile as the corresponding one of \pa,
 only with free intensity factors.%%--2016Feb26:--
 \footnote{Except the width of the narrow \pc, for which we adopt the fitting result
 with it set to be free (cf. Table~3).
 This is because the narrow \pc\ component stands high over the broad component
 and the fitting can be significantly improved by relaxing its width from being tied to that of 
 narrow \pa.}
 %%% Below describing HeI ---- 2016Feb26-----
 The profile of the \hei\,$\lambda10830$ emission line is different from that of \pa, 
 except that its blueshifted cusp is located at a similar blueshifted velocity 
 to the cusps of \pa\ and \hb.
 % LWJ: The line profile of \hei$\lambda10830$ is very different from \pa. Besides the blueshifted 
 %cusp, a large proportion of \heiabs$\lambda10830$ is also blueshifted under the cusp. 
 We understand that \hei$\lambda10830$ is a high-ionization line (with a ionization potential of 24.6~eV),
 and it is well known that high-ionization lines generally have a more complex profile than low-ionization lines
 (e.g., \civ\,$\lambda1549$, \citealp{2011ApJ...738...85W}; see also \citealp{2011ApJ...737...71Z}).
 This is interpreted by the presence of pronounced other components in high-ionization lines---particularly 
 the emission originated from the AGN outflows---in addition to the normal component emitting from the virialized BLR clouds
 that is located at the systematic redshift (see, e.g., \citealp{2011ApJ...737...71Z})
 % LWJ: Compared with lines of low-ionization, broad high-ionization lines may have both a gravitationally 
 % bound, virialized component and a significant outflowing component
 % (e.g., C\,IV, \citealp{2011ApJ...738...85W}).
 To get a convergent fitting for the \hei\ line, 
 we use the profile of the broad \pa\ (namely the two-Gaussian model) as the template to model 
 the virialized broad \hei\ component,
 %% of broad \hei$\lambda10830$ that is originated
 %% from the gas gravitationally bounded in the BLR, 
 a Gaussian to model the NLR-emitted \hei, a Gaussian to the blueshifted cusp, 
 and as many more Gaussians as statistically guaranteed (namely $F-$test probability $\leq$0.05) to account for 
 the remaining flux.
 In the fitting the absorption region is carefully masked. 
 The best-fit model is shown in the lower left panel of Figure~\ref{fig:linefit}
 with a reduced $\chisq =1.15$; the best-fit parameters as listed in Table~3.
 Besides the virialized broad component, the NLR one and the cusp, finally there are two additional Gaussians to account for 
 an extra blueshifted broad component. 
This extra component is blueshifted (by $1360$~\kms), %%please check this number --2016Feb26
which is consistent with the aforementioned outflow interpretation for the profiles of high-ionization broad lines.
Note that such a blueshifting is not a direct identification but merely 
ascribed to the asymmetry of the broad-line \hei$\lambda$10830 profile, which is different from the situation of 
the blueshifted narrow cusp component.
Hereafter when necessary, for the ease of narration we denote this blueshifted broad component with ``outflow$^{\rm B}$'', and 
the blueshifted narrow cusp  ``outflow$^{\rm N}$''.

 %------------------------------------------------------------- Ha + Hb + Hc
 In the optical,
  \ha\ shows a strong broad base and an apparent narrow peak, which are blended 
 with \nii$\lambda\lambda6548,6583$ doublet. The red wing of the broad base is also slightly 
 affected by \sii$\lambda\lambda6716,6731$ doublet.
 \hb, as we stress in the above, shows an apparent narrow cusp blueshifted by $\approx 900$~\kms,
 which is consistent with the extra cuspy components revealed in \pa\ and \hei$\lambda10830$ (see 
 Figure \ref{fig:outflow}).
 This suggests that \ha\ should also have such a blueshifted cuspy component.
 %% 2016Feb26 ---
 Because the Balmer lines are heavily blended with strong \feii\ multiplet emission,
  we fit the continuum-subtracted spectrum (namely, simultaneously fitting Balmer lines $+$ 
  \oiii\ $+$ \nii\ $+$ \sii\ $+$ \feii),
  following the methodology of Dong~\etal\ (2008).
 Specifically, we assume the broad, narrow and blueshifted cuspy components of \hb\ and \hc\ 
 have the same profiles as the respective components of \ha.
 %LWJ: The whole profile of every \oiii$\lambda\lambda4959,5007$ doublet line is 
 %apparently blueshifted and asymmetrical, and we use a double-Gaussian profile to model each and 
 %assuming the doublet line have the same width.
 The \oiii$\lambda\lambda4959,5007$ doublet lines are assumed to have the same profile
 and fixed in separation by their laboratory wavelengths; 
 the same is applied to \nii$\lambda\lambda6583,6548$ doublet lines
 and to \sii$\lambda\lambda6716,6731$ doublet lines.
 %% 2016Feb26 ----
 The flux ratio of the \oiii\ doublet $\lambda5007/\lambda4959$ is fixed to the
 theoretical value of 2.98 \citep[e.g.,][]{2000MNRAS.312..813S,2007MNRAS.374.1181D}; %% check citep, an ir-recognizable character
 the flux ratio of the \nii\ doublet $\lambda6583/\lambda6548$ 
 is fixed to the theoretical value of 2.96 \citep[e.g.,][]{1989Msngr..58...44A,2000MNRAS.312..813S}.
 We use Gaussians to model every components of the above emission lines as we describe in the above for the NIR lines,
 starting with one Gaussian and adding in more if the fit can be improved significantly 
 according to the $F-$test. 
 The best-fit model turns out that 
 two Gaussians are used for the broad component of the Hydrogen Balmer lines:
 one for the narrow component, and one for the blueshifted cusp.
 Two Gaussians are used for every line of the \oiii\ doublet and one for all the other aforementioned narrow lines.
 The \feii\ multiplet emission is modeled with  
 the two separate sets of analytic templates of Dong~\etal\ (2008), one for the broad-line \feii\
 system and the other for the narrow-line system, constructed from measurements
 of I~Zw~1 by \citet{2004A&A...417..515V}. Within each system, the respective 
 set of \feii\ lines is assumed to have no relative velocity shifts and the 
 same relative strengths as those in I~Zw~1. We assume that the broad and narrow 
 \feii\ lines have the same profiles as the broad and narrow \hb, respectively;
 see Dong~\etal\ (2008, 2011) for the detail and justification. 
 A first-order polynomial is adopted to fine-tune the local continuum of the \ha\ $+$ \nii\ $+$ \sii\ region 
 and the \hb\ $+$ \hc\ region, respectively.
 The best-fit model is presented in Figure~\ref{fig:linefit}, 
 with a reduced $\chisq$ of 2.34 in the \hb\ $+$ \hc\ region and a reduced $\chisq$ of 1.46 
 in \ha\ $+$ \nii\ $+$ \sii\ region.  
 The somehow large reduced \chisq\ in the \hb\ $+$ \hc\ region 
 is mainly due to the excess emission in the red wing of \hb,
 the so-called ``red-shelf'' commonly seen in type-1 AGNs and
 has been discussed in the literature \citep[e.g.,][]{1985PASP...97..734M,2002A&A...384..826V}.
 It is probably the residual of \feii\ multiplet 42 ($\lambda4924, 5018, 5169$),
 or broad \hei$\lambda\lambda4922,5016$ lines
 \citep[see][]{2002A&A...384..826V}, or just the mis-match between \ha\ and \hb.
Since it is irrelevant to the components of interest in this work, we do not discuss it further.
We also fit \oii\,$\lambda3727$ line, which is well isolated and easily fitted with two Gaussians (see Figure~5).

 The measured line parameters are listed in Table~\ref{tab:emline}.
 The extinction of the broad, narrow, and blueshifted cuspy components of the Balmer lines
 can be derived from the observed Balmer decrement \ha/\hb. 
 The intrinsic value of broad-line \ha/\hb\ is 3.06 with a standard deviation of 
 0.03 dex \citet{2008MNRAS.383..581D}.
 A value of 3.1 is generally adopted for the intrinsic narrow-line \ha/\hb\ in AGN
 \citep{1983ApJ...269L..37H,1984PASP...96..393G}.
 The intrinsic \pa/\hb\ of AGNs is very close to the Case-B value 0.34
 \citep{1984PASP...96..393G}.
 In Table \ref{tab:decrement}, we list the observed \ha/\hb\ and \pa/\hb\ as well
 as $E_{\rm B-V}$ assuming the extinction curve of the LMC ($R_{\rm V} = 2.6$).
 The extinctions to the broad components and to the blueshifted
 cusps are similar, while the narrow components suffer much lager
 extinction, indicating that the NLR could be more dust-obscured.
  %% ????????extinction????????????????????
  %, which indicates the narrow line region is more dusty than 
  %the broad line region and outflow gas. 

\subsubsection{Absorption Lines \label{subsec:measureabs}}

  \thisobj\ shows \heiabs$\lambda\lambda3889, 10830$ and \nai~D BALs, which are 
  generally deemed  to be caused by AGN outflows. 
  To analyze the absorbed intensities, we need to first identify the 
  pre-absorption AGN spectrum and normalize the data with it.
  %We use the best fit model to normalize the absorption line spectrum.
  We first subtract the best-fit narrow emission lines and the starlight
  component from the observed spectrum before normalization, as the
  absorption gas does not cover the NLR in all well studied BALs.
  There are three components left in the observed spectrum: the power-law continuum,
  the broad emission lines, and the blackbody (hot dust) continuum.
  \heiabs\ absorption lines might be normalized in different ways depending on
  whether or not the absorbing gas covers the torus and/or the BLR.
  We notice that the \heiabs$\lambda10830$ absorption trough shows a flat bottom,
  which indicates that the \heiabs$\lambda10830$ line is saturated and the residual
  fluxes should be zero in those pixels.
  Meanwhile, we find the residual fluxes at the line centroid of \heiabs$\lambda10830$ is 
  close to zero after subtracting the starlight continuum (see the middle panel in 
  Figure~\ref{fig:conti}). 
  Again, we check carefully the decomposition of the continuum.
  %%%-- In English, the normal order is: adv. after the verb, unless 
  %%% you want to stress the adv word.
  %%% 
  The fractions of the power law and starlight in the optical are determined by
  the decomposition of images, which should be reliable.
  The fitted blackbody continuum, which originates from the hot dust of the torus, is the
  dominant emission in the $K$ band and the $WISE W1$ band,
  which also accords with the expectation.
  Therefore, we conclude that the absorbing gas is likely exterior to the torus.
  
\begin{figure}[htbp]
   \center
  \includegraphics[width=3.2in]{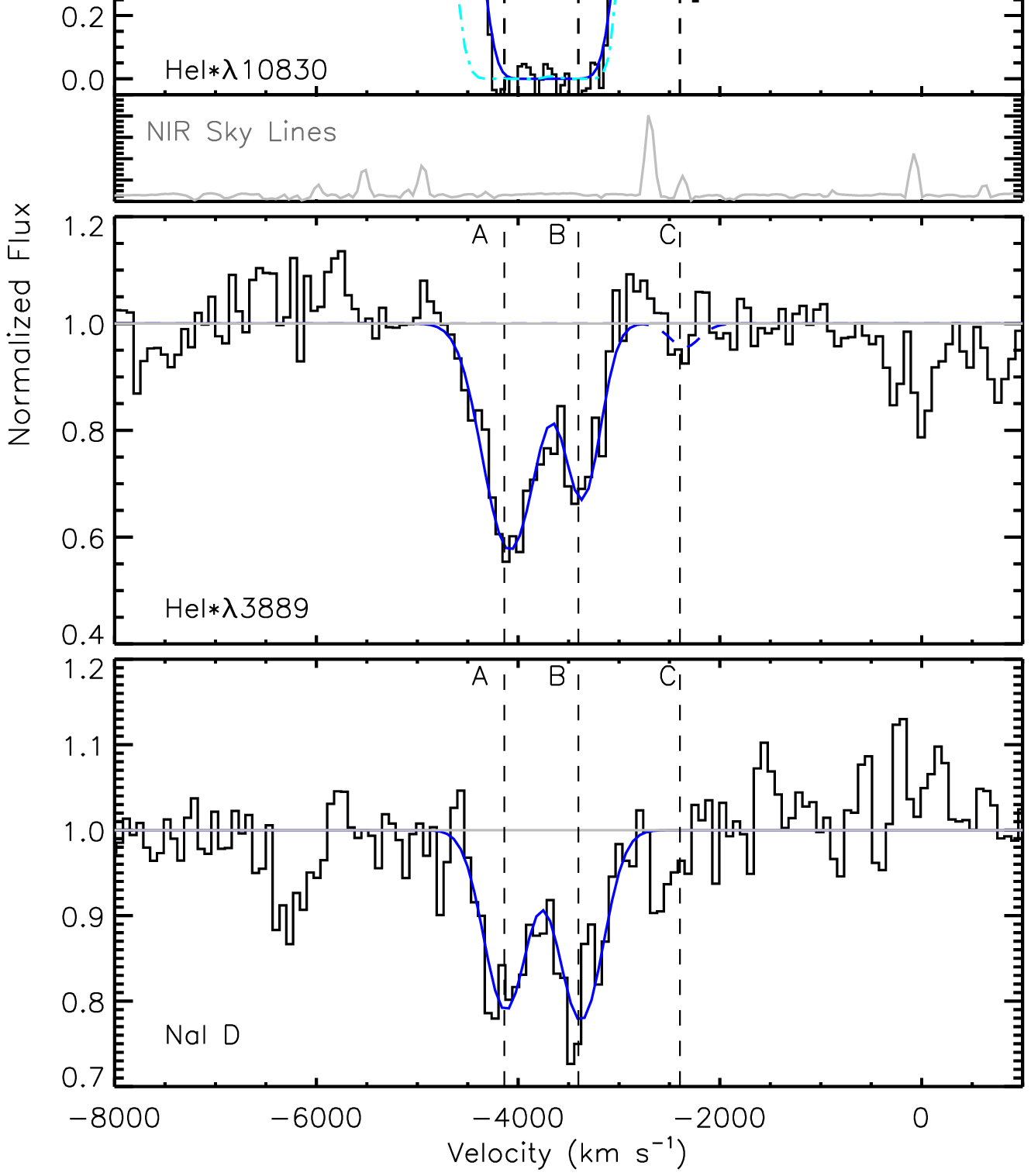}\includegraphics[width=3.2in]{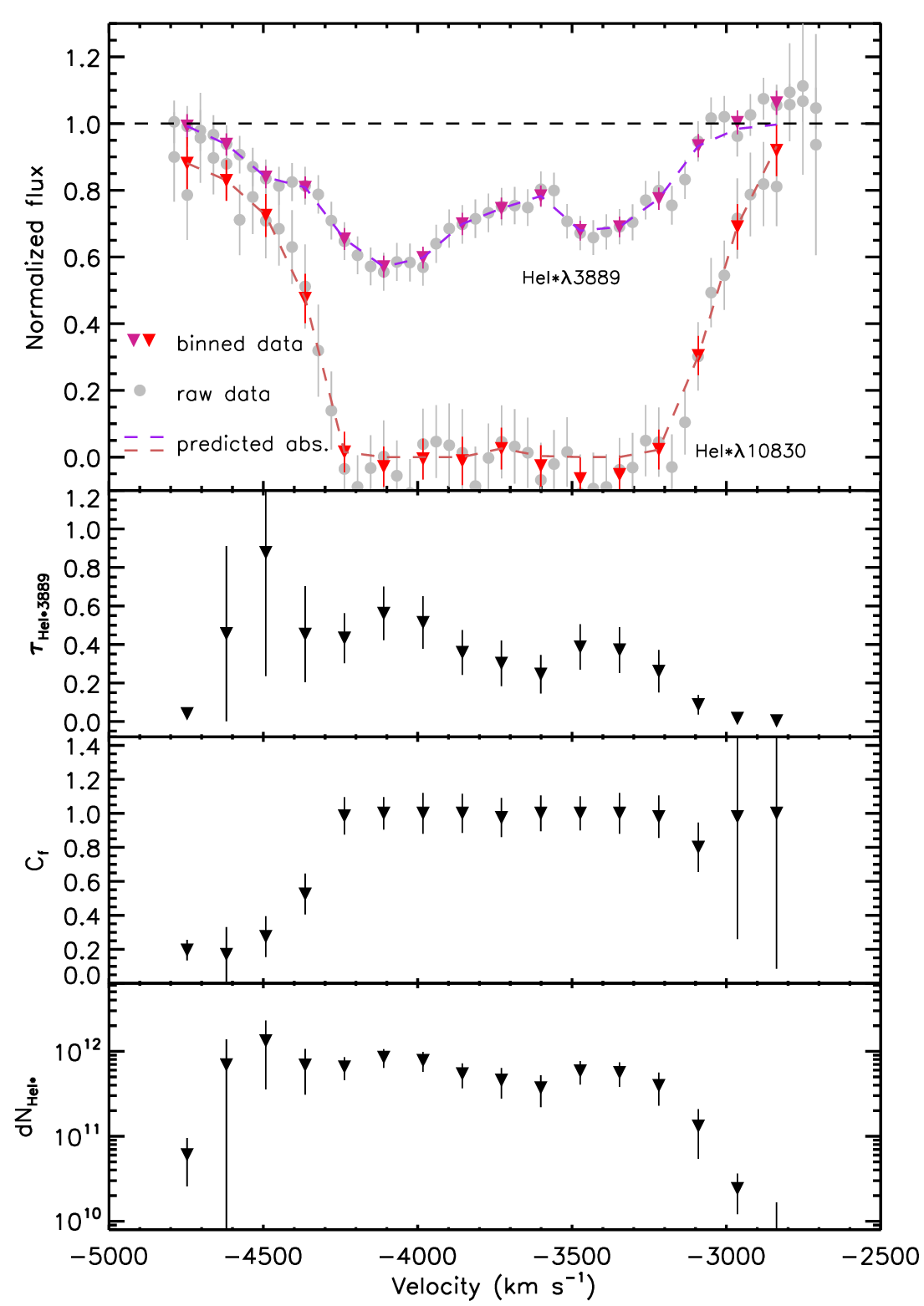}
  \caption{\footnotesize
    {\it Left}: demonstration of the absorption troughs in \heiabs$\lambda\lambda10830,3889$
    and \nai~D of \thisobj, and the best fittings (in blue) by Voigt profile.
    The cyan lines in the upper panel show the predicted \heiabs$\lambda10830$
    profile derived from the \heiabs$\lambda3889$ absorption line under
    the fully coverage assumption.
    {\it Right}: calculation results (top panel) of the \heiabs\ absorption lines
    of \thisobj\ using the partial-coverage model.
    The top panel shows the observed profiles (the gray dots) with $\pm 1\sigma$ error
    bars of \heiabs$\lambda\lambda10830,3889$ the profiles binned by 3 pixels
    (the colored triangles) and the predicted profile from calculation (the dashed lines).
    The model parameters (the optical depth $\tau_{\rm \heiabs 3889}$, 
    covering fraction ($C_{\rm f}$), and column density ($dN_{\rm \heiabs}$)
    of the outflow) as a functions of velocity are shown in the other three panels. 
    \label{fig:abs}}
\end{figure}

  The left panels of Figure~\ref{fig:abs} demonstrate the absorption profiles of
  \heiabs$\lambda\lambda3889,10830$ and \nai~D absorption lines in velocity space.
  On the SDSS spectrum, the \heiabs$\lambda3889$ absorption is clearly detected. 
  It splits into two absorption troughs, including a larger one near -4100 \kms, and 
  a second one near -3400 \kms, and totally covering the velocity ranges -5000 -- -2800 \kms\  
  (trough A and B in Figure \ref{fig:abs}).
  The strong \heiabs$\lambda10830$ absorption line is identified on the TripleSpec spectrum.
  It separates into two major troughs spreading from $v \sim -5000$ \kms to $v \sim
  -2000$ \kms\ (trough A+B, C in Figure \ref{fig:abs}).
  \heiabs$\lambda3889$ and \heiabs$\lambda10830$ are transitions from the same lower level,
  so they should have the same velocity profile theoretically.
  The strongest trough, which corresponds to the trough A and B of \heiabs$\lambda3889$,
  is obvious saturated, since the bottom of the absorption line is flat. 
  This can be naturally explained by the large optical depths ratio of 
  \heiabs$\lambda10830$ and \heiabs$\lambda3889$. 
  The \nai~D$\lambda\lambda5890,5896$ profile also two principal components, the velocities 
  of which correspond to the absorption troughs A and B of \heiabs$\lambda3889$.
  The appearance of neutral Sodium absorption lines suggests the large column densities of 
  the absorption gas and the neutral Sodium exists deep in the clouds, otherwise neutral Sodium
  will be easily ionized.
  The third \heiabs$\lambda10830$ trough (trough C) has a weak counterpart of \heiabs$\lambda3889$
  and \nai~D.

  We use the Voigt profile \citep{1938ApJ....88..508H,1984ApJ...278..486C} to fit 
  these absorption troughs, which are shown in Figure~\ref{fig:abs}. 
  The Voigt profile is implemented with the program x\_voigt in the XIDL package.%
  \footnote{http://www.ucolick.org/~xavier/IDL/}
  \heiabs$\lambda\lambda3889,10830$ are the strongest two transitions from the 
  metastable state to the 2p, 3p states, and the $\tau$ ratio 
  ($\varpropto \lambda~f_{\rm ik} N_{\rm ion}$) of \heiabs$\lambda\lambda10830,3889$ 
  is 23.5:1 \citep[their Table 2]{2015ApJS..217...11L}.
  If the lines are not saturated and the absorbers fully cover the source, the normalized 
  flux $R_{10830} = R_{3889}^{23.5}$.
  The cyan dotted lines in the upper left panel of Figure~$\ref{fig:abs}$ shows the 
  \heiabs$\lambda10830$ absorption profile predicted from the \heiabs$\lambda3889$ 
  absorption trough under the full-coverage assumption.
  The red wing of the observed \heiabs$\lambda10830$ absorption line fits 
  the predicted profile well, while the blue wing of the observed profile is evidently
  different from the predicted.
  This suggests a full-coverage situation for the absorbing gas of low outflowing speed  
  and a partial coverage for the gas of high speed.

  Based on the above inference, we try to estimate the covering factor of the outflow
  gas assuming a simple partial-coverage model, where the observed normalized spectrum 
  can be expressed as follows: 
    \begin{equation} 
          R = (1 - C_{f}(v)) + C_{f}(v)e^{-\tau(v)} .
    \end{equation}
  Here the $\tau$ ratio of \heiabs$\lambda\lambda10830,3889$ is 23.5.
  Although the spectral resolutions of the SDSS and TripleSpec spectrum are not 
  high enough for us to study the velocity structure of absorption 
  trough in detail, the tendency and the mean of $C_{\rm f}$ are reliable.
  We bin the spectrum by three pixels and perform the calculation and analysis 
  using the binned data.
  Following the methodology of \citet{2011ApJ...728...94L}, we derive the covering
  fraction ($C_{\rm f}$), the optical depth ($\tau$) and the column density of \heiabs,
  as a function of velocity (see right panels of Figure \ref{fig:abs}).
  the covering factor of outflow with $v < 4300$~\kms\
  is close to 1.0, contrasting with the high speed outflow with $v > 4300$~\kms.
  The average covering factor of component A is 0.82, and
  log~$N_{\rm \heiabs} (\rm cm^{-2})= 15.01\pm0.16$.
  Assuming \nai~D absorption lines have the same covering factor with \heiabs\ absorption 
  lines, then we get the ionic column density of \nai\ is 
  log~$N_{\rm HaI} (\rm cm^{-2})= 13.31\pm0.21$.

%% 2015June28eve ---
   
\subsection{Analysis of the $Spitzer$ spectrum \label{subsec:mirspec}}

\begin{figure}[tbp]
  \center
  \includegraphics[width=6.4in]{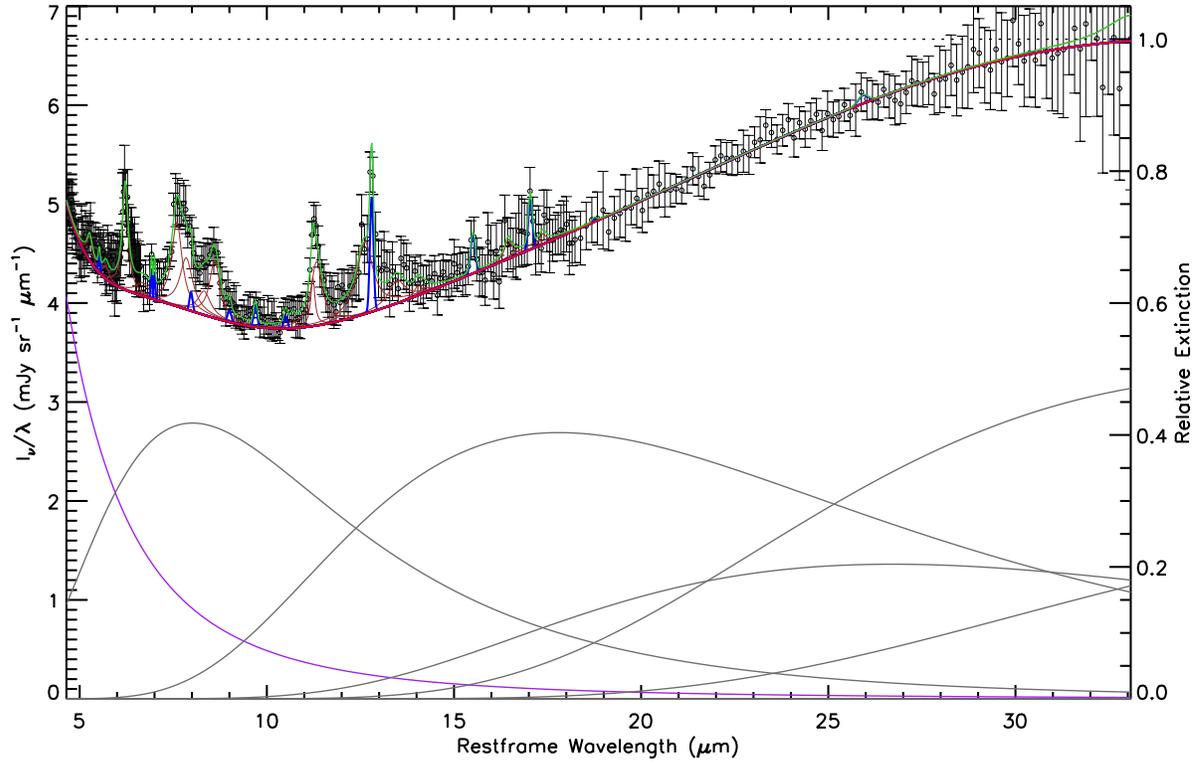}
  \caption{Best-fit decomposition of \thisobj\ from 5 to 33~$\micron$ (rest frame) from PAHFIT.
           Details of the model parameters can be found in \citet{2007ApJ...656..770S}.
       The black circles with error bars show the observed data.
       The gray lines represent the thermal dust continuum components, the purple line
       shows the stellar continuum, and the red line shows the total (dust + stellar)
       continuum.
       The brown lines represent the PAH features and the blue lines show the atomic and
       molecular lines.
       All components are diminished by the fully mixed extinction, indicated by the dotted
       black lines, with the axis at the right.
       The green line is the full fitted model, plotted over the observed flux intensities and
       uncertainties.\label{fig:mir}}
\end{figure}

 %% added by Wenjuan on July 13th
  The MIR spectrum of \thisobj\ shows significant PAH emission features and a 
  steeply rising continuum toward long wavelength end (see Figure \ref{fig:mir}).
    We measured the \rev{apparent strength 
    (namely the apparent optical depth)} of the 9.7$\micron$ silicate feature
  following the definition by \citet{2007ApJ...654L..49S},
  \begin{equation}
    S_{\rm sil} = \text{ln} \frac{f_{\rm obs(9.7 \micron)}}{f_{{\rm cont}(9.7 \micron)}}  ~,
  \end{equation}
  where $f_{\rm obs(9.7 \micron)}$ is the observed flux density at 9.7$\micron$
  and $f_{\rm cont(9.7 \micron)}$ is the continuum flux density at 9.7$\micron$.
  Following \citet{2007ApJ...654L..49S}, we estimate $f_{\rm cont(9.7 \micron)}$ 
  from a power-law interpolation over 5.5--14.0\micron.
  It gives $S_{\rm sil} =  0.01 \pm 0.03$ (1$\sigma$),
  indicating (almost) neither silicate emission nor absorption.
  It is interesting that the silicate absorption is not present at all, as
  the analysis in the UV and optical bands demonstrates instead that \thisobj\ is
  dust-obscured. In the diagnostic diagram of EW\,(PAH 6.2~\micron) vs. 9.7 \micron\ silicate
  strength as shown in Figure~1 of \citet{2007ApJ...654L..49S},
  galaxies are located mainly around two branches closely: a diagonal  and a horizontal.
  \thisobj\ belongs to the horizontal branch in the 1A class
  (EW(PAH~6.2\micron) = 0.050$\pm$0.003, see Table~\ref{tab:mirpar}).
  Interestingly, Mrk~231 also belongs to this class but with a larger 9.7$\micron$
  silicate absorption strength ($S_{\rm sil} \sim -0.65$) as well as a smaller 
  PAH 6.2$\micron$ EW ($\sim 0.01$).
  As discussed by \citet{2007ApJ...654L..49S}, the two distinct branches reflect
  likely the differences in the spatial distribution of the nuclear dust.
  Galaxies in the horizontal branch may have clumpy dust distributions, which
  produce only shallow absorption features.
  In \thisobj\ as we analyzed above, the narrow emission lines suffer larger
  extinction than the broad emission lines
  and the young stellar component is obscured more seriously
  than the AGN continuum and broad/narrow emission lines.
  This implies that
  the dust in the nuclear region is clumpy and patchy; i.e.,
  the nuclear region is not fully enshrouded by dust.
  Meanwhile, there are significant \neii~12.8\micron\ and
  \neiii~15.6\micron\ emission lines, but no higher-ionization lines such as
  \nev~14.3\micron\ and \ovi~25.89\micron\ in the MIR spectrum;
  this may indicates that the AGN contributes insignificantly to the MIR emission
  \citep{2007ApJ...667..149F}.

   %% using PAHfit code ---
  We use the PAHFIT spectral decomposition code (v1.2)\,%%
  \footnote{http://tir.astro.utoledo.edu/jdsmith/research/pahfit.php}
  \citep{2007ApJ...656..770S} to fit the spectrum as a sum of dust attenuated 
  starlight continuum, thermal dust continuum, PAH features, and emission lines.
  As the above analysis suggests that the AGN contribution in the MIR flux is low,
  we adopt the fully mixed extinction geometry.
  Figure~\ref{fig:mir} shows the MIR spectrum and the best-fit decomposition with
  a reduced $\chisq = 1.16$.
  The $\tau_{9.7}$ is fitted to be $\sim 0$.
  The fluxes of the main MIR components derived form the PAHFIT are listed in 
  Table~\ref{tab:mirpar}.
  %%% -- 2016Feb26: ----
  The flux errors are given by PAHFIT; see Smith~\etal\ (2007) for the detail.
  The errors of the EWs are estimated according to error propagation formula,
  where the uncertainties of continuum is estimated as follows.
  For the spectral region of every emission line feature, 
  we calculate the residual between the raw spectrum and the best-fit (continuum + line)
  and then the standard deviation of the residual in this region is taken as the 1-$\sigma$ error of the continuum placement. 
  The total PAH Luminosity Log~$L_{\rm PAH}$ (erg s$^{-1}$) = 43.64$\pm0.02$, which is 
  $\sim$1.2\% of total IR luminosity of this source.
  %Moreover, the hardness indicator [NeIII]/[NeII] of \thisobj\ is $0.375$,
  %and the ratio $L(\rm PAH 7.7\micron)$/$L(\rm PAH 11.3\micron) = 2.45$.
  %\thisobj\ is in the transitional region of H~II-dominated nuclei and AGN
  %types \citep[see their Figure~14]{2007ApJ...656..770S}, which is also consistent
  %with the discussion above.
  We estimate star formation rate from the PAH features using the relation of 
  \citet{2007ApJ...667..149F},
  SFR(\msun, yr$^{-1}) = 1.18\times10^{-41}~L\,(\rm PAH~6.2\micron+\rm PAH~11.2\micron)$, and
  the errors of the SFR derived using this equation are of order $\sim$ 50\% for individual objects.
  According to our measurements, we get the SFR$_{\rm PAH} = 141\pm71~\msun \rm yr^{-1}$.
  As a double check, we also estimate the SFR from the 24$\micron$ emission. \rev{Adopting the 
  relation for galaxies with $L_{\rm IR} \geq 10^{11}\lsun$ of \citet{2009ApJ...692..556R},
  the error of the SFR derived from which is whitin 0.2 dex,
  we get SFR$_{24\micron} \sim 143^{+84}_{-53} ~\msun \rm yr^{-1}$}.
  The SFR values estimated from the PAH, 24$\micron$ emissions and the total IR luminosity are well consistent,
  and they are also consistent with the upper limit of the SFR estimated from the IR luminosity.
  Hence, we adopt the SFR estimated from PAH in this paper.

%% 2015June28: ----
 \section{Results}
 
  \subsection{Central Black Hole}

   With the measured luminosity and line width of the broad emission lines, 
   we can estimate the mass of the central BH using the commonly used virial 
   mass estimators.
   We use the broad \ha\ based mass formalism given by
   \citet[their Eq.~6]{2011ApJ...739...28X}, which is based on
   \citet{2005ApJ...630..122G,2007ApJ...670...92G} but incorporates the recently updated
   relation between BLR size and AGN luminosity calibrated by \citet{2009ApJ...697..160B}.
   The broad-\ha\ luminosity is corrected for the broad-line extinction
   using the LMC extinction curve, resulting $L_{\rm \ha} = 1.5 \times 10^{43}$ \ergs.
   Together with FWHM(\ha) $= 2955$ \kms, the central Black hole mass is 
   estimated to be $M_{\rm BH} = 7.94 \times 10^{7} \msun$.

   %% a new paragraph ---
   We calculate monochromatic continuum luminosity $\lambda L_{\lambda}(5100\AA)$
   at 5100 \AA\ from the best-fit power-law component (see \S\ref{subsec:continuum}).
   The best-fit $E_{\rm B-V}$ value of the power-law component is 0.42 (assuming the LMC
   extinction curve) and the extinction corrected luminosity
   $\lambda L_{\lambda}(5100\AA) = 2.47 \times 10^{44}$\ergs.
   As a check, we also estimate the $\lambda L_{\lambda}(5100\AA)$ from the \ha\ flux
   \citep{2005ApJ...630..122G}, which gives $\lambda L_{\lambda}(5100\AA) = 2.49 \times 10^{44}$ \ergs,
   fairly consistent with the above one obtained from the best-fit power law.
   \rev{Then we calculate the bolometric luminosity using the conversion 
   $L_{\rm bol} = 0.75 \times 4.89 + 0.91 \lambda L_{\lambda}(5100\AA)$ \citep{2012MNRAS.422..478R}, 
   which gives $L_{\rm bol} = 1.45 \times10^{45}$~\ergs.
   The corresponding Eddington ratio is thus $L_{\rm bol}/L_{\rm Edd} = 0.15$.
   Based on the bolometric luminosity, the amount of mass being accreted is estimated 
   as follows: 
   \begin{equation}
     \dot{M}_{\rm acc} = \frac{L_{\rm bol}}{\eta c^{2}} = 0.26 \msun \rm yr^{-1}  ~,
   \end{equation}
 where we assumed an accretion efficiency $\eta$ of 0.1, and $c$ is the speed of light.}

%% $3.2 ---
   \subsection{Host Galaxy}
  
    %% the dominating old stellar population ---
   The 2D decomposition of the SDSS images (\S2.3) yields that the 
   host galaxy is an early-type galaxy with S\'{e}rsic index $n = 4$ 
   and $M_{\rm r} = -22.07$~mag.
   We use the $K$-band luminosity of the starlight to calculate the stellar mass
   of \thisobj, which is relatively insensitive to dust absorption and to stellar 
   population age. \citet{2013MNRAS.430.2715I} provide a calibration of mass-to-light
   ratios against galaxy colors, and we here adopt their Table 3 relation for 
   log $M/L_{\rm Ks}$ = 1.055 ($B-V$) - 1.066 with a scatter of $\pm 0.13$ dex. The $B$, $V$
   magnitude and $L_{\rm Ks}$ of \thisobj\ is calculated by convolving the decomposed 
   stellar component (see \ref{subsec:continuum}) with the Jonson $B$ band and $V$ band response
   curve. We get $B-V = 0.63$ and $L_{\rm Ks} = 2.4 \times 10^{44}$ \ergs, 
   the stellar mass estimated from $L_{\rm Ks}$ is $M_{\rm host} = (1.8 \pm 0.54) \times 10^{11} \msun$. 
   The error estimate accounts for the uncertainty of decomposition of starlight in 
   $K$ band (see \S \ref{subsec:continuum}) and the uncertainty of $M/L_{\rm Ks}$.
   As a check, we also use V band luminosity to estimate the stellar mass. 
   The UV-NIR continuum decomposition (\S \ref{subsec:continuum}) shows that the galaxy is 
   dominated by an old stellar population with an age of $\approx 9$ Gyr, which corresponds to a 
   mass-to-light ratio of $M/L_{\rm V} \approx$ 5.2 \msun/\lsunv\ \citep{2003MNRAS.344.1000B}.
   The $V$-band luminosity of the old stellar population is calculated by convolving the decomposed
   old stellar component with Jonson $V$-band response curve, which gives $L_{\rm V} = 9.24 \times 10^{43}$ \ergs.
   Thus, the stellar mass is $M_{\rm host} \approx 1.34 \times10^{11}\msun$, which is 
   basically consistent with stellar mass estimated from $K$-band luminosity.

  % The optical continuum fitting result (\S2.4) shows that the galaxy is
  % dominated by an old stellar population with an age of 
  % %%$\sim 9$ Gyr,
  % $\approx 9$  %%2016Feb25
  % which corresponds to a mass-to-light ratio
  % $M/L_{\rm V} \approx$ 5.2 \msun/\lsunv\  \citep{2003MNRAS.344.1000B}.
  % We calculate its V-band luminosity of this old stellar population by
  % convolving its best-fit spectrum with the Jonson V-band response curve,
  % giving $L_{\rm V} = 9.24 \times 10^{43}$ \ergs.
  % The stellar mass of this component is thus $M_{\rm host} = 1.34 \times10^{11}\msun$.
%% pls update the numbers. 
%% pls get the stellar mass using the K-band luminosity. --2016Feb25.

   %% the reddened young stellar population ---2015June30,night ---
   Both the decomposition of the UV to NIR spectrum and the analysis of its SED and MIR spectrum
   reveal that there is a heavily obscured young stellar component (see \S \ref{subsec:continuum})
   relating to the recent violent (obscured) star formation activities.
   This stellar component is heavily obscured in the UV and optical bands.
   We cannot see any sign of star formation activities from the optical image either.
   According to the decomposition of the optical continuum, this component accounts
   for $\lesssim 5$\% of the total continuum emission at 5100\AA.
   The best-fit result of the spectral decomposition gives the extinction of this 
   component is E$_{\rm B-V} =~2.2$ (see \S \ref{subsec:continuum}).
   This is why we can only reliably infer the presence of this young stellar component from the PAH 
   emission in the $Spitzer$ spectrum and the high FIR luminosity.
   %According to the SFR estimated from PAH emission and the relation
   %$SFR(\msun yr^{-1}) = 4.5\times 10^{-44} L_{\rm IR}$ (\ergs) \citep{1998ARA&A..36..189K},
   %the IR luminosity from the star formation should be 3.1$\times10^{45}$ \ergs,
   %which accounts for $\sim$88\% of the total observed IR luminosity.

   %%- to create a new paragraph here: 2015Feb25 -----
   It is interesting to note the fact that
   the AGN narrow emission lines suffer more dust obscuration than the AGN broad lines,
   and the young stellar component suffer more than the two.
   %the extinction of the NLR is larger
   %than BLR, and the extinction of young SSP is larger than the nuclear region.
   This may tell us some clues of the spatial distribution of the dust. The AGN NLR is probably
   related to the heavily obscured \hii\ region.
   Next we investigate the situations of the observed narrow emission lines in all the available bands,
   which are in principle powered by both starburst and AGN.
   \rev{The optical narrow emission lines are observed to have the following line ratios:
   log~(\hb/\oiii$\lambda5007$) $=~0.86\pm0.03$,%%2016Feb25
   \footnote{\label{footnote:oiiibpt} As the whole profile of \oiii$\lambda5007$ is blueshifted 
   (see Figure~\ref{fig:outflow}), 
   which is not the case of the other unshifted narrow emission 
   lines (e.g., \hb, \ha).
   We only integrate fluxes within $-500$---500 \kms\ corresponding to the \hb\ narrow lines.}
   %%% end of the footnote---
   log~(\nii$\lambda6583$/\ha) $=~0.03\pm0.01$, log~(\sii$\lambda\lambda6716,6731$/\ha) $=~-0.23\pm0.01$,
   and log~(\oi$\lambda6300$/\ha) $=~-0.85\pm0.04$}.
   According to the BPT diagnostics diagrams 
   \citep{1981PASP...93....5B,2003MNRAS.346.1055K,2006MNRAS.372..961K},
   \thisobj\ is classified as a purely Seyfert galaxy, indicating that these observed optical
   line fluxes are mainly powered by the AGN.
   Then how are the NIR narrow emission lines? As we know, the NIR narrow emission lines 
   should be less affected by dust obscuration, and so the observed narrow \pa\ could be powered 
   by both the AGN and the star formation considerably. 
   Thus, a conservative starburst-powered \pa\ emission can be estimated as follows.
   We take (reasonably) the narrow \hb\ to be totally from the AGN NLR, and correct it for the 
   nuclear dust reddening by the narrow-line Balmer decrement (see Table~\ref{tab:decrement}),
   giving the unreddened narrow \hb\ flux of $1.3\pm0.32 \times 10^{-14}$~\flux.
   In AGN NLR environment, the line ratio \pa/\hb\ is close to the Case-B value for all conditions
   and the value is $\sim$~0.34 \citep{1984PASP...96..393G}. 
   Thus, the AGN-powered narrow \pa\ flux (unreddened) can be estimated from the unreddened narrow 
   \hb\ flux, giving $4.4\pm1.1 \times 10^{-15}$~\ergs~cm$^{-2}$.
   Similarly, we estimate the total unreddened narrow \pa\ flux by applying the same dust extinction 
   as the narrow \hb, which is certainly a conservative estimate (lower limit) 
   of the \pa\ flux (since the star-formation region is obscured more seriously than the AGN NLR);
   this yields the total unreddened narrow \pa\ $> 7.3\pm0.3 \times 10^{-15}$ \ergs~cm$^{-2}$ \AA$^{-1}$ 
   (both the AGN and starburst powered).
   So the AGN-powered is at most 61\% of the total narrow \pa\ flux.
   That is, more than 39\% of the narrow \pa\ is powered by starburst.

   %%% 2015July1, noon : ------
    %%% below M-sigma_* ----
   With interest, we try to investigate the relationship between the galactic 
   bulge and the central BH.
   As we cannot %% "can not" is wrong. --2016Feb25 ---
   measure the stellar velocity dispersion from the optical spectra, 
   we use instead the velocity dispersion of the line-emitting gas in the NLR
   as traced by the low-ionization [S~II]$\lambda\lambda6716,6731$ lines as a surrogate
   \citep{2005ApJ...627..721G,2007ApJ...667L..33K}.
   For $\sigma_{\star} \simeq \sigma_{\rm [SII]} = 178.8\kms$,
   the $M_{\rm BH}$--$\sigma_{\star}$ relation for early-type galaxies of
   \citet[their Table 2]{2013ApJ...764..184M} predicts $M_{\rm BH} = 1.37 \times 10^{8}\msun$,
   which is 1.6 times the virial BH mass based on the broad \ha\ line.
   Now we take the S\'{e}rsic $n=4$ component (\S2.3) as the galactic bulge, the
   $M_{\rm BH}$--$L_{\rm bulge}$ relation for early-type galaxies of
   \citet[their Table 2]{2013ApJ...764..184M} predicts $M_{\rm BH} = 3.47 \times 10^{8}\msun$,
   which is $\sim$ 4.4 times the virial estimate based on the broad \ha\ line.
   Considering the large uncertainties associated with these quantities (both methodologically and
   statistically), at this point we can only say that both the \mbh--$\sigma_{\star}$
   and \mbh--$L_{\rm bulge}$ relationships of the nearby inactive galaxies seem to hold in this object.

%%% $3.3:
   \subsection{The companion galaxy and the galactic ring \label{subsec:ring}}

\begin{figure}[htbp] 
      \centering
        \begin{minipage}[c]{0.37\textwidth}
        \centering
        \includegraphics[width=2.4in]{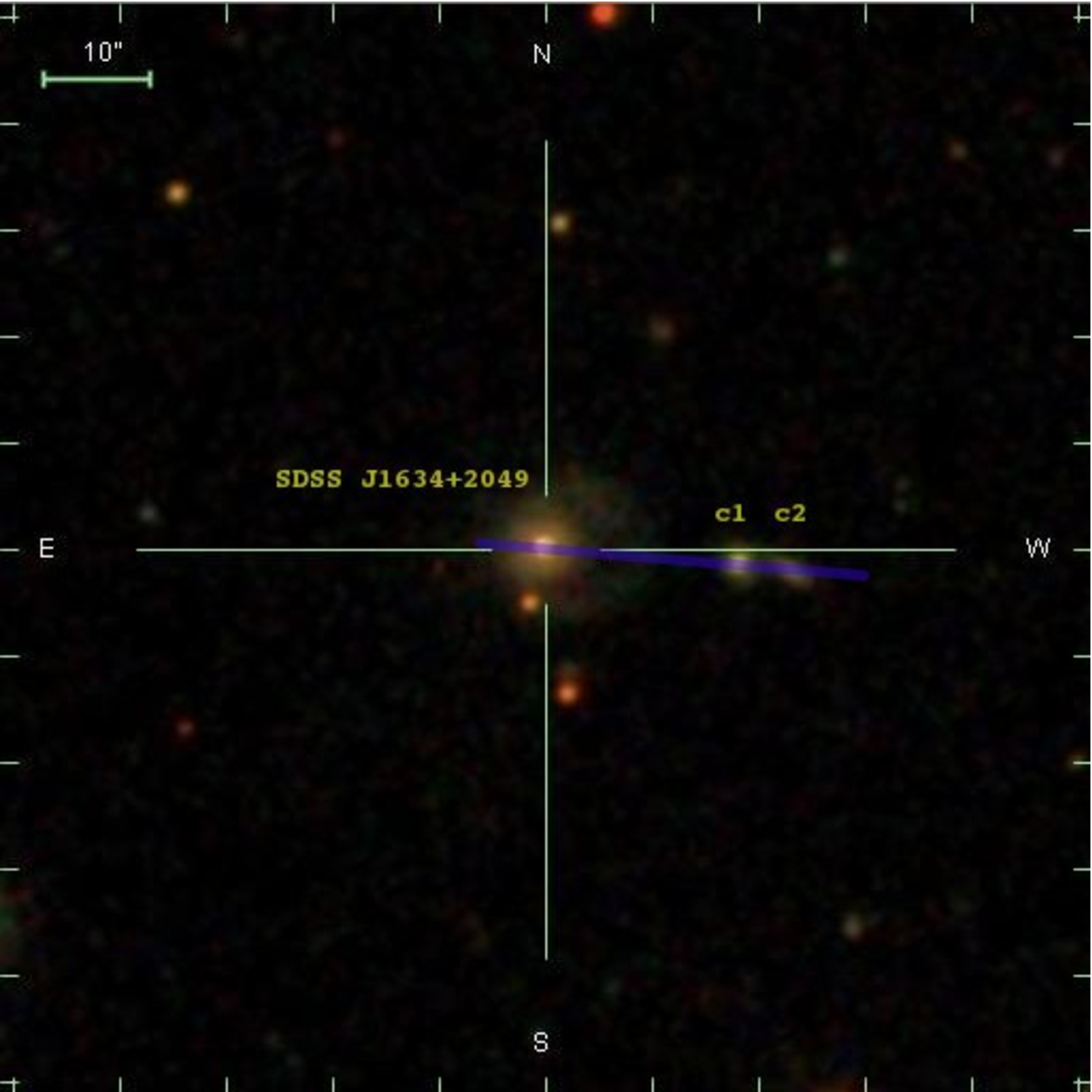}
        \end{minipage}%
    %\hspace{0.1in}
    \begin{minipage}[c]{0.62\textwidth}
            \centering
        \includegraphics[width=3.8in]{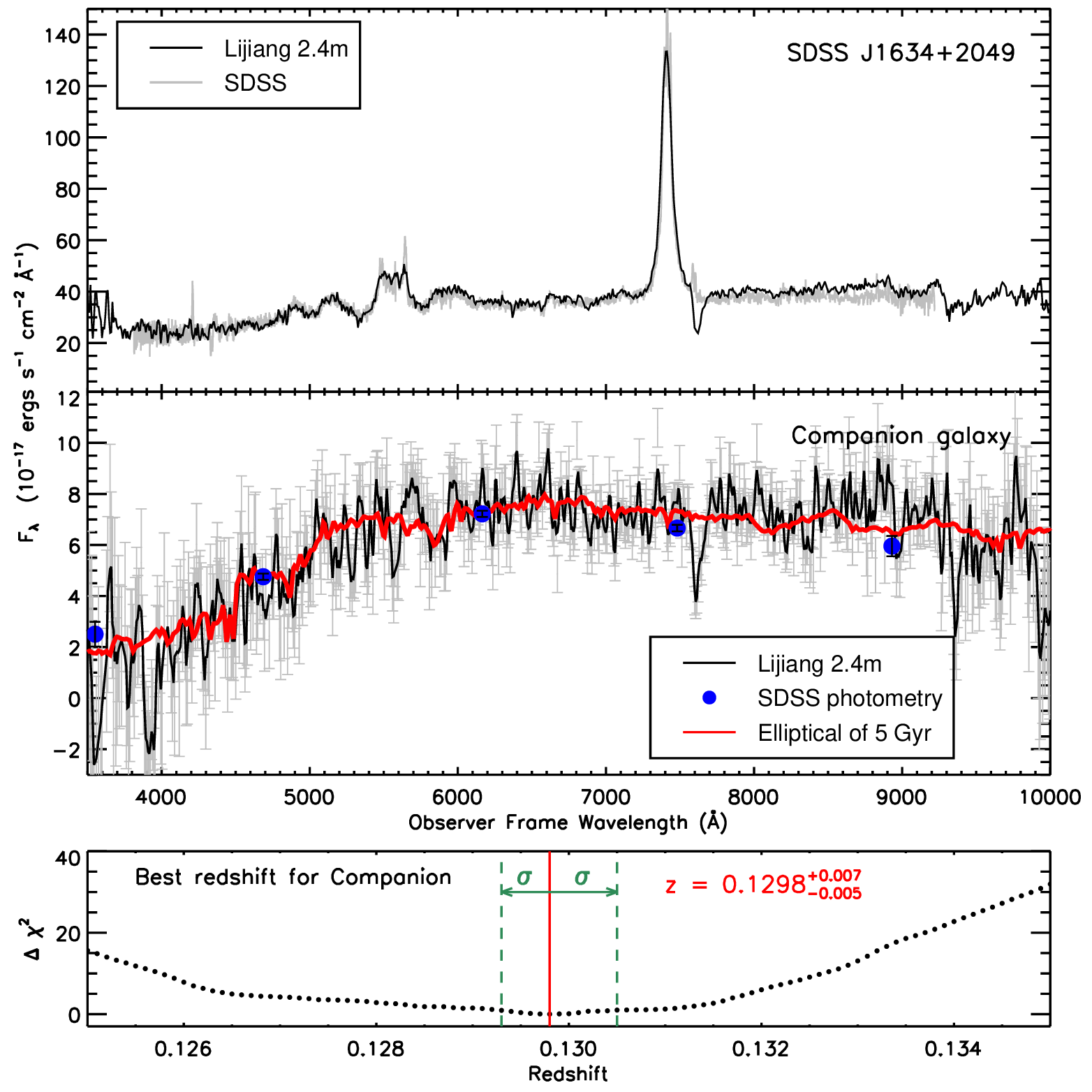}
    \end{minipage}
    \caption{\footnotesize {\it Left:} SDSS image of \thisobj\ and its companion galaxy, C1.
    The PA = 86\arcdeg\ slit position is displayed by a blue line.
    {\it Right:} the upper panel shows the Lijiang 2.4m spectrum in the observer frame of \thisobj\
    SDSS spectrum of \thisobj\ (gray line in upper panel) is displayed for a comparison.
    The middle panel displays the Lijiang 2.4m spectrum of the companion galaxy C1.
    SDSS photometry of C1 are also plotted on the observed spectrum as a check.
    The red line represents a template of an elliptical of 5 Gyr, which is obtained
    from \citep{2007ApJ...663...81P}.
    The lower panel shows the $\Delta \chi^{2}$ of the grid search of the redshift of C1.
    The best redshift is $z=0.1298^{+0.007}_{-0.005}$. \label{fig:companiongal}}
    \end{figure}

   %% revised by Xufen Wu July 9th
   There are two small faint galaxies to the west of of the ring 
   %%on the W side
   (C1 and C2 in Figure~\ref{fig:companiongal}).
   The projected distances from the center of \thisobj\ to the
   centers of C1 and C2 are 18\arcsec.5 and 19\arcsec.1, corresponding to
   35.7 kpc and 44.0 kpc, respectively.
   We spectroscopically observed the C1 and C2 galaxies using YFOSC
   mounted on Lijiang GMG 2.4m telescope.
   The right panel of Figure \ref{fig:companiongal} shows the obtained
   spectra of \thisobj\ and C1.
   The C2 galaxy is too faint to get an effective spectrum.
   In the right top panel of Figure~\ref{fig:companiongal}, the spectra of the 
   Lijiang 2.4m telescope and of the SDSS are compared to make sure that the 
   wavelength calibration is reliable, which is key to determine the 
   redshift of C1.
   The spectrum of C1 galaxy shows characteristics of an early-type
   galaxy (ETG) of old stellar population, with visible stellar
   absorption features and no emission lines. By comparing the C1
   spectrum with the template spectra in the SWIRE template library 
   \citep{2007ApJ...663...81P}, a best-matched template of elliptical 
   galaxy of 5 Gyr old is picked up.
   Because of no strong emission lines, we perform a grid search of redshift 
   for the redshift of C1. The searching procedure is as follows.
   The redshift grids are set to be $0.11 < z < 0.15$.
   In every redshift grid, the template spectrum is brought to the observer frame,
   and is reddened with the Milky Way extinction along the direction of C1;
   then we fit the observed C1 spectrum with the reddened template and calculate
   the $\chi^{2}$.
   In this way we obtain the curve of $\chi^{2}$ with redshift, as shown in the 
   bottom panel of Figure~\ref{fig:companiongal}.
   The redshift value corresponding to the smallest $\chi^{2}$ is taken as the 
   best-fit redshift of the C1 galaxy, which is 0.1276 with the 1-$\sigma$ error of 
   0.0004 ($\Delta \chi^{2} = 1$).
   Recalling the redshift of \thisobj\ being $z = 0.1293$, the redshifts of the 
   \thisobj\ and C1 implies that the two galaxies form a collisional system with
   a line of sight (LOS) velocity difference of $\approx 451$ \kms.
   We also calculate the $V$ band luminosity for C1 by convolving the spectrum with
   Jonson $V$ band filter, yielding $L_{\rm V} = 7.8 \times 10^{42}$ \ergs.
   Though the best-matched template for C1 shows a stellar age of 5 Gyr, the real
   stellar age of C1 is hard to determine with the low signal-to-noise ratio(S/N) spectrum.
   The colors ($g-r$, $r-i$) of C1 from the photometry are (1.03, 0.34), which are even
   redder than that of \thisobj. It is likely that the stellar age of C1 is older than
   9~Gyr. 
   %However, considering the uncertainties of photometry of this faint object, we 
   %can only safely say that the stellar age of C1 is older than 5 Gyr.
   To be conservative, we argue the stellar age of C1 should be older than 5 Gyr.
   For a stellar population of solar metallicity with an ages of 5 Gyr and 12 Gyr,
   the mass-to-light ratios $M/L_{\rm V}$ are between 3.36 and 6.68 \msun/\lsunv\ \citep{2003MNRAS.344.1000B}.
   Thus the stellar mass of C1 $M_{\rm c1}$ is estimated to within
   $7.3 \times 10^{9}$ -- $1.45\times 10^{10}$\msun.

   Figure~\ref{fig:2dimage} shows that \thisobj\  is an ETG (an elliptical or
   a S0 galaxy) with a ring structure.
   The off-centered ring is an ellipse in the sky, with its major and minor axes
   being $\sim$ 14\arcsec.0 (32.3~kpc) and $\sim$ 12.\arcsec7 (29.4~kpc), respectively. 
   In \S\ref{subsec:sdssimage} we measured the quantities of the ring as well as the main 
   body (namely the S\'{e}rsic) of the host galaxy from the SDSS images (see Table~\ref{tab:galfit}).
   The colors ($g-r$, $r-i$) of the S\'{e}rsic component are (0.93, 0.5), and the ring, (0.94, 0.48).
   The similarity of the colors between the two components
   suggests that the ring may have the same stellar population as
   %% 2016Feb26 --
   the S\'{e}rsic component, (averagely) aging $\approx9$--12~Gyr.
   Taking the above implication, the ring has $L_{\rm V,ring} = 1.9 \times 10^{43}$ \ergs,
   and stellar mass $M_{\rm ring} = 2.9 \times10^{10}\msun$. 
   %% -- 2016Feb26, pls update the above numbers ---

   There are several theories proposed to explain the formation and evolution
   of the ring galaxies: (1) head-on collisions between a disk galaxy and a intruder
   with a mass of at least one tenth of the disk galaxy
   \citep[e.g.,][]{1976ApJ...209..382L, 1977ApJ...212..616T,1978IAUS...79..109T,2008MNRAS.383.1223M,
   2012MNRAS.420.1158M,2012MNRAS.423..543S}; 
   (2) Lindblad resonances which forms a smooth ring with a central nucleus and with 
   the absence of companion galaxies \citep{1999AJ....117..778B}; and (3) an accretion 
   scenario which forms the polar ring galaxies \citep{2003A&A...401..817B}.
   The ring around \thisobj\ and observation on the nearby galaxy, C1, suggest a head-on 
   collision scenario for the formation of the ring.  
   Besides, the same color of the stars in the ring and the host galaxy confirms 
   such a formation history.
   The ring around \thisobj\ is offset to the galactic center (the nucleus), which resembles
   the collisional RN class of galactic rings proposed by \citet{1976ApJ...208..650T}.
   Numerical simulations showed that the ring structure can be created in the stellar, 
   the cold gas, or both contents of galactic disk by the radial propagation of a density 
   wave which is formed in the collision
   \citep[e.g.,][]{1976ApJ...209..382L,1977ApJ...212..616T,1978IAUS...79..109T,2008MNRAS.383.1223M,
   2012MNRAS.420.1158M}.
   Numerical simulations also showed that an off-center collision can produce
   the offset of the central nucleus and the elliptical rings
   \citep[e.g.][]{1976ApJ...209..382L,2012MNRAS.420.1158M,2015ApJ...805...32W}.
   The inclination angle of the collisional parent galaxies are related to the
   ellipticity of a galactic ring \citep[Eq.~7]{2015ApJ...805...32W}.
   The ellipticity of the ring around \thisobj\ is $\epsilon \equiv 1-b/a\approx 0.09$,
   where $a$ and $b$ are the major and minor axes of the projected ring.
   The ellipticity of the ring is small, so that the off-center collision and
   inclination effects are small.
   Besides, the impact parameter (i.e., the minimal distance between the bullet and
   the disk galaxy) sensitively affects the morphology of the ring \citep{1978IAUS...79..109T}.
   The ring is more lopsided and the nucleus is more offset when the impact parameter
   is larger. The ring structure disappears when the impact parameter is too large,
   and spiral structure forms instead. The ring structure is clearly around the
   \thisobj, and it is offset. Thus we may infer that \thisobj\ and C1
   had experienced an off-centered collision with a small impact parameter.
   %the collision does not exactly along the symmetry axis of the \thisobj, meanwhile the impact parameter
   %is small enough to form the ring.
   %Taking all above, an approximation that the collision is almost along the symmetry
   %axis of \thisobj\  is reasonable.

    %% 2015July2 noon, starts here: -------------
    \subsection{Determining the physical condition of the outflows \label{sec:outflow}}
    
   As described in \S2.4.3, strong blueshifted absorption troughs show in the optical 
   and NIR spectra, indicating the presence of strong outflows.
   The decomposition of the emission line profiles (\S \ref{subsec:emissionline}) also
   indicates that the presence of outflows in emission as revealed by the
   blueshifted line components of almost all the observed emission lines.
   In this subsection, we analyze and determine the physical properties of
   both the absorption line and emission line outflows
   using photoionization synthesis
   code {\it Cloudy} \citep[c13.03;][]{1998PASP..110..761F}.

  \subsubsection{The absorption line outflow \label{subsec:abscloudy}}

   \thisobj\ shows \heiabs$\lambda\lambda3889,10830$ and \nai~D absorption 
   lines in its NUV, optical and NIR spectra. 
   The physical conditions of the absorbers are quite different 
   to generate \heiabs\ and \nai~D absorption lines.
   The metastable $2s$ state in the helium triplet, \heiabs\, is populated via
   recombination of He$^{+}$ ions, which is ionized by photons with energies of
   $h\nu \geq 24.56$ eV.
   Therefore, \heiabs is a high-ionization line and its column density
   ($N_{\rm \heiabs}$) mainly grows in the very front of hydrogen ionization
   front and stops growing behind it
   \citep[e.g.,][]{2001ApJ....546..140A,2015ApJ...800...56J,2015ApJS..217...11L}.
   Instead, \nai~D absorption line is produced by neutral Sodium, the potential 
   of which is only 5.14~eV, and is easily destroyed by the hard AGN continuum;
   therefore, this line is rare to detect in quasar spectra.
   It can only exist where dust is present to shield neutral sodium from the intense 
   UV ($\geq 5.14$~eV) radiation filed of the AGN that would otherwise photoionize it 
   to Na$^{+}$.
   In \S\ref{subsec:measureabs}, we get the column densities of \heiabs\ and \nai\ for
   the major absorption trough A+B (see Figure \ref{fig:abs}),  
   log$N_{\rm \heiabs} (\rm cm^{-2})= 15.01\pm0.16$, log$N_{\rm HaI} (\rm cm^{-2})= 13.31\pm0.21$.
   Low-ionized \caii~H\,\&\,K absorption lines usually present with \nai~D lines.
   \caii~H\,\&\,K  absorption lines arise from Ca$^{+}$ ions, which is ionized by photons
   with energies of $h\nu \geq 6.11$ eV and destroyed by photons with energies of
   $h\nu \geq 11.8$ eV.
   We find no apparent \caii~H\,\&\,K absorption lines in the NUV spectrum, which is
   probably because calcium element is depleted into the dust grains.
   We estimate the upper limit of column densities \caii\ by shifting \nai~D absorption
   profile to \caii~H wavelength, and use the profile as a template to fit the \caii~H region.
   The upper limit of column density of \caii\ is log~$N_{\rm \caii}$(cm$^{-2}$) = 12.9.
   All of the above indicate that the dust should be considered in the photoionization models.
   In addition, we also notice there are no apparent Balmer-line absorption lines in the
   spectrum of \thisobj, and we get the upper limit of column densities of hydrogen
   n=2 is log $N_{\rm H (n = 2)}$ = 12.6 cm$^{-2}$.
   This suggests the electron densities of the absorbing gas cannot be higher than 
   10$^{8}$ cm$^{-2}$ \citep{2011ApJ...728...94L,2012RAA....12..369J}.

   Here we present a simple model calculated by {\it Cloudy}
   \citep[c13.03;][]{1998PASP..110..761F} to explore the conditions required to
   generate the measured \heiabs\ and \nai\ column densities.
   We start by considering a gas slab illuminated by a quasar with a density of 
   $n_{\rm H}$ and a total column density of $N_{\rm H}$.
   The SED incident on the outflowing gas has important consequences for the
   ionization and thermal structures within the outflow. Here we adopt the
   UV-soft SED, which is regarded more realistic for radio-quiet quasars than
   the other available SEDs provided by {\it Cloudy} (see the detailed discussion
   in \S4.2 of \citet{2010ApJ...709..611D})
   The UV-soft SED we adopted here is a superposition of a blackbody ``big bump''
   and power laws, and is set to be the default parameters given in the {\it Hazy}
   document of {\it Cloudy} where $T=150,000$~K, $\alpha_{\rm ox} = -1.4$,
   $\alpha_{\rm uv} = -0.5$, $\alpha_{x} = -1$ and the UV bump peaks at around 1 Ryd.

   %% - to start a new paragraph -- about handling the dust ----
   The above analysis shows the absorption material is a mixture of dust grains 
   and gas, and thus the dust-to-gas ratio and the depletion of various elements from 
   the gas phase into dust should be taken into account in the models.
   The data of \thisobj\ in hand favor the presence of dust of the LMC extinction 
   type, which certainly needs UV spectroscopic observations to confirm.
   In the following model calculations, we use the {\it Cloudy}'s built-in model 
   of ISM grains and assume the total (dust$+$gas) abundance of the absorption 
   material to be the solar abundance.
   So the gas-phase abundance changes with the dust-to-gas ratio. 
   %%-- 2015July3 , afternoon, after wenjuan's nap ----
   %%---
   As the {\it Cloudy} software does not account for the conservations of the mass
   and abundance of the elements that are both in the gas phase and depleted in
   the dust grains, trying to keep the total abundances of various element to be
   the solar, in the iterated {\it Cloudy} calculations we manually set
   the gas-phase abundance according to the dust content we add.
   The {\it Cloudy}'s built-in grain model of ISM dust that incorporates C, Si, O, 
   Mg and Fe elements.
   Regarding the Ca and Na elements, we adopt the built-in scheme of dust depletion,
   with their gas-phase depletion factors being 10$^{-4}$ and 0.2, respectively, 
   which are the default recipe of {\it Cloudy}.%
   \footnote{The full description of the depletion scheme is given in 
     \S7.9.2 of the {\it Cloudy} 13.03 manual {\it Hazy} and references therein.}
   Besides element abundances, the dust-to-gas ratio ($A_{\rm V}/N_{\rm H}$)
   is another key parameter in the {\it Cloudy} model involving dust.
   According to the measured narrow-line and broad-line extinction, we simply adopt 
   $A_{\rm V} = 2$ for the dust in the cloud slab.
   The total column density $N_{\rm H}$ of the slab can be mutually constrained by
   comparing the measured \nai~D, \heiabs, and the non-detections (upper limits) of \caii\
   and Balmer absorption lines with the Cloudy simulations in the parameters spaces.

\begin{figure}[htbp] 
    \center
    \includegraphics[width=6.4in]{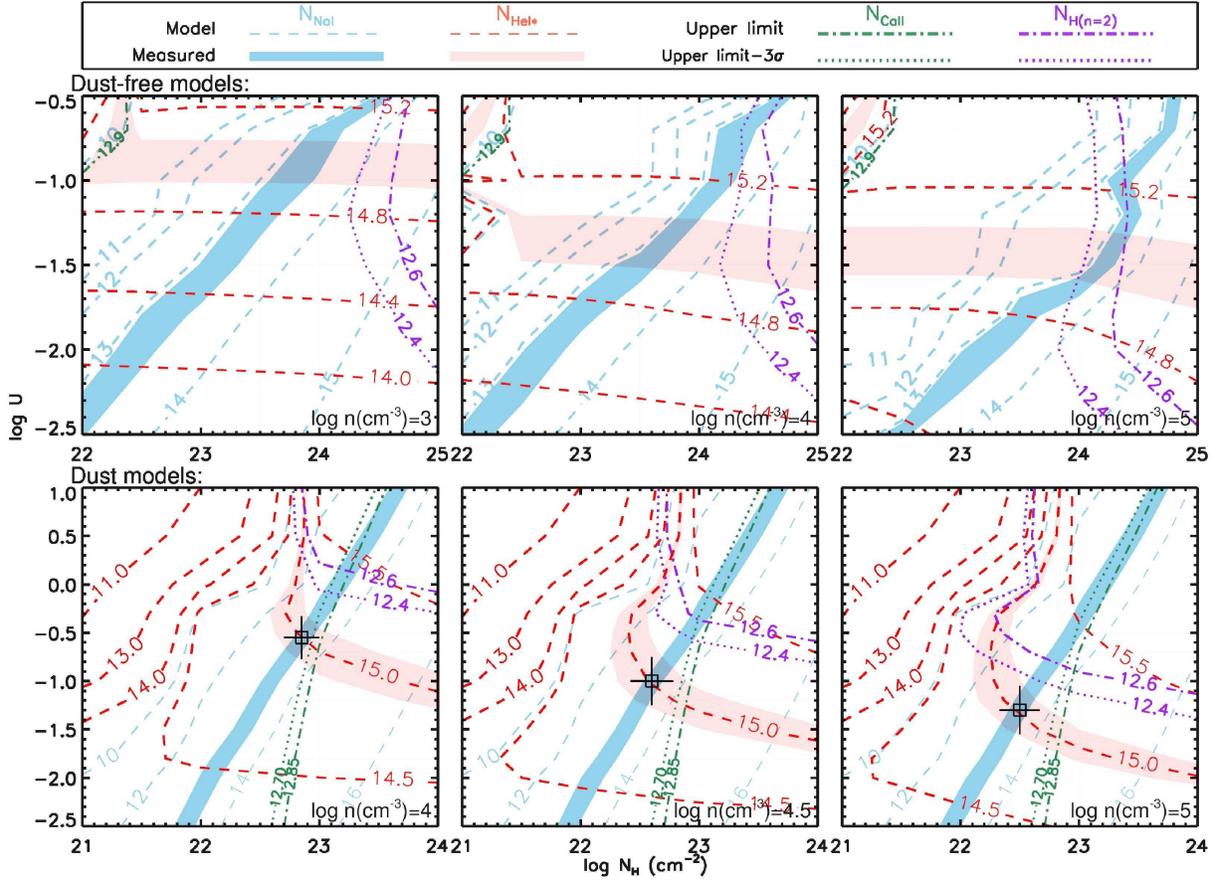}
    \caption{\footnotesize Photoionization models of the absorption outflow in \thisobj\
      assuming solar abundances. 
      The upper three panels show results of the dust-free models, and the lower three
      panels show the models added the effects of dust grains mixed in the gas slab.
      Red and blue dashed lines represent the contours of ionic column densities of
      \heiabs\ and \nai, respectively.
      The red and blue shaded regions represent the locus of points ($N_{\rm H}$, $U$)
      able to reproduce the observed $N_{\rm \heiabs}$ and $N_{\rm \nai}$ with 1-$\sigma$ error
      respectively.
      The purple and the green dotted-dashed lines represent a upper limit on the ionic column densities
      of H($n$=2) and \caii. 
      Correspondingly, the purple and the green doted lines represent the upper limit minus 3 $\sigma$ 
      error of H($n$=2) and \caii, which indicate the confidence interval of the upper limit and 
      also the decreasing direction of the ionic column density.
      The best models are marked by an open square with the systemic on the solution. 
      \label{fig:cloudyabs}}
   \end{figure}

   %% -start a new paragraph here: 2015July5 evening-----
   We start the simulations first with the dust-free baseline models to get the initial
   total column density $N_{\rm H}$ (the most probable value; see below),
   and then we feed {\it Cloudy} with the dust-to-gas ratio, $A_{\rm V}/N_{\rm H}$
   calculated by this $N_{\rm H}$ and get a new $N_{\rm H}$.
   Then we use this new $N_{\rm H}$ into {\it Cloudy} and start a new iteration.
   The simulations will be stopped when the $N_{\rm H}$ value is convergent.
   During every iteration, the {\it Cloudy} simulations are run over the grids of
   the following parameter space: ionization parameter  $-2.5 \leq$ log U $\leq 1.0$,
   hydrogen density $3 \leq$ log $n_{\rm H}$ (cm$^{-3}$)$\leq 7$, and the stop column
   density $21 \leq$ log N$_{\rm H}$(cm$^{-2}$) $\leq 25$
   (see Liu~\etal\ 2015 for the detail of the {\it Cloudy} modeling).
   By comparing in the $N_{\rm H}$--$U$ plane the measured ionic column densities with 
   those simulated by {\it Cloudy}, we get the allowed parameter intervals for 
   $N_{\rm H}$ and $U$ as constrained mutually by the various absorption lines.
   The details are illustrated in Figure~\ref{fig:cloudyabs}.

    %%%%%%----------- The following of this paragraph is for Wenjuan to revise.  ------------%%%%%%%%%%%%%
    %%%%%%---           dxb, 2015July5 night ---------------------%%%%
   In Figure \ref{fig:cloudyabs}, the upper three panels display dust-free 
   models and the lower three panels display the best models added the 
   effects of dust grains mixed in the gas slab.
   In dust-free models, the upper limit of $N_{\rm \caii}$ is not considered, 
   since the measured value is much smaller than the models predict due to
   the heavily dust depletion.
   We can see that dust-free models with log 
   $n(\rm H)$ (cm$^{-3}$) = 3 -- 5 are in accord with the measurements for 
   this object.
   The dust-free models give the initial $N_{\rm H}$ (overlap region) and 
   the initial dust-to-gas ratio ($A_{\rm V}/N_{\rm H}$) for the following 
   iterative calculation of dust models.
   The initial value of $N_{\rm H}$ are around $10^{24}$ cm$^{-2}$, and the 
   initial $A_{\rm V}/N_{\rm H} \sim 2 \times 10^{-24}$ cm$^{-2}$ which is 
   1.7\% of the ratio of LMC, $A_{\rm V}/N_{\rm H} = 1.2 \times 10^{-22}$ cm$^{2}$ 
   \citep{2001ApJ...548..296W}.
   The lower three panels show the convergent solutions (black open squares) 
   of dust models.
   The suitable solutions (log $n_{\rm H}$, log$N_{\rm H}$, log $U$) for the 
   absorption line outflow of \thisobj\ are (4, 22.85$\pm0.15$, -0.55$\pm0.22$), 
   (4.5, 22.6$\pm0.18$, -1.0$\pm0.25$) and (5, 22.5$\pm0.17$, -1.3$\pm0.25$).

  %%-- 2015June3 morning, starts: ---
   \subsubsection{Emission-line outflows \label{subsec:cloudyel}}

    As Figure \ref{fig:outflow} demonstrates, \thisobj\ also shows emission line
    outflows as revealed by blueshifted hydrogen Balmer and Paschen lines,
    \heiabs$\lambda\lambda10830,5876$, \oiii$\lambda\lambda4959,5007$,
    \oii$\lambda3727$ and \neiii$\lambda3869$ lines,
    particularly by the blueshifted, well-separated cuspy components.
    This narrow (FWHM $\lesssim 650$~\kms corrected for the instrumental broadening) 
    cusp exists only in the recombination lines, invisible in any forbidden emission 
    lines.
    It is blueshifted by $900$~\kms\ with respect to the system redshift
    consistently in the hydrogen Balmer and Paschen lines and \hei\ lines;
    particularly, in the \hb\ and \hei$\lambda10830$ this component is well-separated 
    from the peaks of the normal NLR- and BLR-originated components,
    obviously not an artifact of the line profile decomposition.
    Interestingly to note, in light of its absence in forbidden lines, 
    this line component should be  originated from the dense part of the outflows.
    A second blueshifted line component, is present in all high-ionization
    emission lines such as \hei$\lambda\lambda5876,10830$,
    \oiii$\lambda4959,5007$ and \neiii$\lambda3869$, with a similar
    asymmetric profile and similar best-fit FWHM of $\approx 1600$\kms.
    This component is  much stronger in flux and relatively broader than
    the cuspy component. Although in \hei$\lambda10830$
    it is not so well-separated from the normal NLR and BLR components,
    this component manifests itself well in the forbidden lines 
    \oiii$\lambda\lambda4959,5007$ and \neiii$\lambda3869$ with the 
    whole emission line profile being blueshifted, since there are no 
    other line components in these forbidden lines.
    %%%
    The low-ionization forbidden line \oii$\lambda3727$ also shows an obvious 
    bulk-blueshifted component, in addition to the normal NLR-originated component
    sit just at the system redshift that exists in all low-ionization, forbidden emission lines
    (e.g., \nii\ and \sii); this bulk-blueshifted component of  \oii$\lambda3727$
    has a smaller blueshift than in the aforementioned high-ionization lines.
    %%
    %Note that, unlike recombination lines, all forbidden lines do not show obvious spikes 
    %at $\sim$ -900 \kms.
    %Taken all above, we can infer that 
    Besides, we can infer that this second blueshifted component (denoted as the broad outflow)
    should be originated from the less dense part of the outflows, with density lower than
    10$^{6}$ cm$^{-3}$.\footnote{The critical densities ($n_{\rm crit}$) of 
    \oiii$\lambda\lambda4959,5007$ and \neiii$\lambda3869$ are 
    $\sim 7.0\times10^{5}$ cm$^{-3}$ and $\sim 9.7\times10^{6}$ cm$^{-3}$ respectively.}
    %%
    %% --the following is move to the above-----
    %In contrast, the narrow blueshifted component is not present in ionization forbidden 
    %lines, and this suggests the outflow gas of this component have higher density than 
    %that of the broad blueshift component.

    %% 2015July3 night : ---

\begin{figure}[htbp] 
     \center
     \includegraphics[width=6.4in]{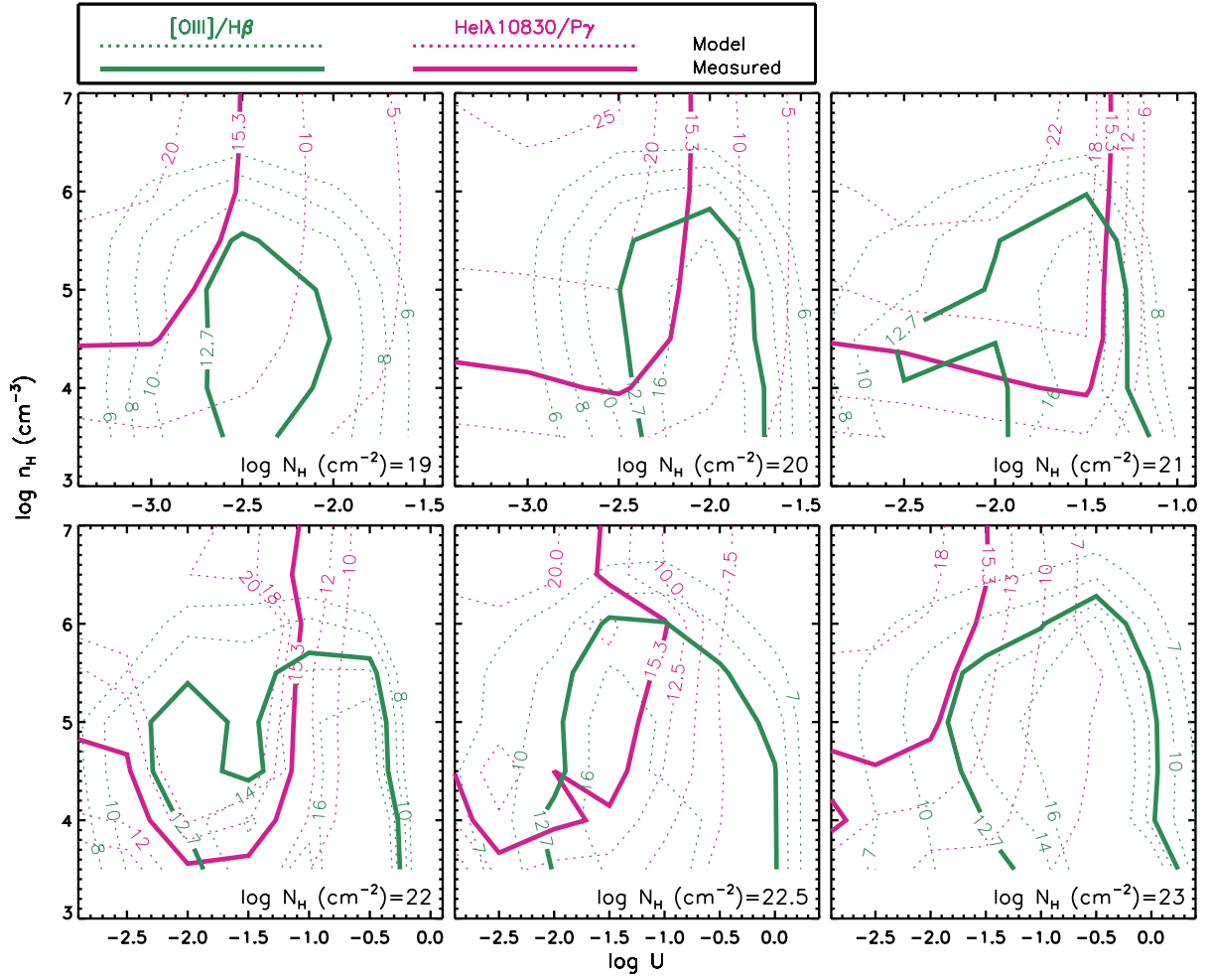}
     \caption{\footnotesize Photoionization models of the emission line outflow in \thisobj\ assuming 
       solar abundances and dust free.
       Different panels are for models with different $N_{\rm H}$.
       The green and violet dashed lines show the contours of line ratios of \oiii$\lambda5007$/\hb\
       and \hei$\lambda10830$/\pc\ in ($U$, $n_{\rm H}$), respectively.
       The green and violet solid lines represent the locus of the measured upper limit for line ratios 
       of \oiii$\lambda5007$/\hb\ and \hei$\lambda10830$/\pc, respectively.
       The closed region surrounded by the green and the violet solid lines are the possible parameter space for 
       the emission line outflow of \thisobj. \label{fig:cloudyel}}
   \end{figure}

\begin{figure}[htbp] 
      \center
     \includegraphics[width=6.4in]{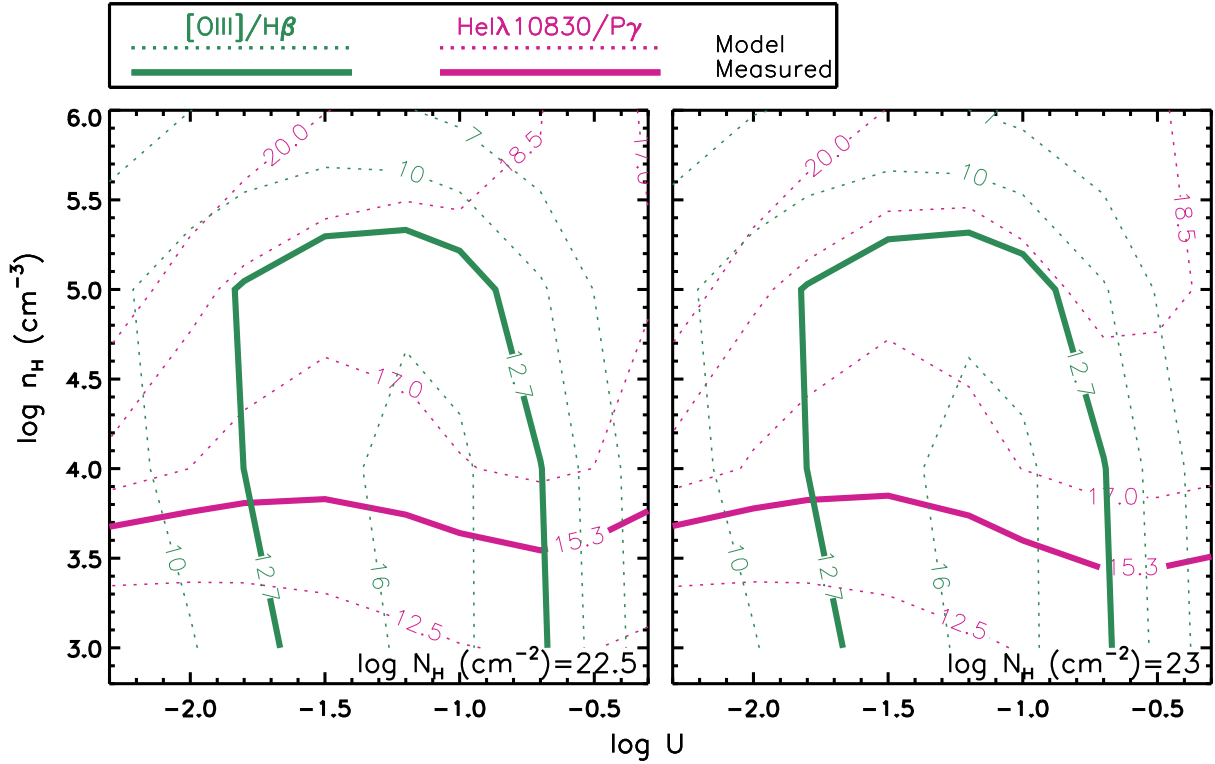}
     \caption{ \footnotesize Dust models of the emission line outflow in \thisobj. 
       The emission line ratios are directly extracted from the best dust models of 
       absorption line outflow. The symbols are the same as Figure \ref{fig:cloudyel}.
       \label{fig:cloudydustel}}
   \end{figure}

    The dynamics of the outflowing gas of this object should be complex. Here we only consider
    about the strong, blueshifted broad component of the emission lines, which is much stronger than the cuspy narrow
    component.
    To investigate the physical condition for the emission line outflow gas, we use the
    {\it Cloudy} simulations calculated above, and then confront  these models with the measured line
    ratios to determine $N_{\rm H}$, $n_{\rm H}$ and $U$.
    As we demonstrate above, both the continuum and emission lines of \thisobj\ are heavily
    reddened by dust. So it is better for us to use the line ratios of adjacent lines to minimize the
    effect of reddening.
    Besides, the difference of ionization potential of the two lines should be
    large in order for them to probe different zone in a gas cloud. 
    Thus, we use the line ratios \oiii$\lambda5007$/\hb\ and \hei$\lambda10830$/\pc\ here.
    Here the flux of \oiii\ is measured by subtracting
    the flux within $\pm$ 500 \kms\ (cf. Footnote~\ref{footnote:oiiibpt}) 
    from the whole \oiii$\lambda5007$ profile, i.e., only the flux blueward of $-$500~\kms;
    the flux of
    \hei$\lambda10830$ is the broad blueshifted component (see Table~3).
    %%% introducing your fluxes used in this part. --- Now the treatment of Hb, Pc. --- 
    We do not detect the broad blueshifted component in hydrogen emission lines,
    which may be weak and concealed in the best-fit blueshifted narrow component;
    so we take the blueshifted narrow components of \hb\ and \pc\ as the upper limits here.
    Therefore, the lower limits of  \oiii$\lambda5007$/\hb\ and \hei$\lambda10830$/\pc\ are estimated
    to be 12.75$\pm1.12$ and 15.28$\pm3.44$. 
    On the modeling part,
    we first extract the simulated \oiii$\lambda5007$, \hei$\lambda10830$, \hb\ and \pc\ fluxes from 
    the dust-free {\it Cloudy} simulations and then compute the line ratios 
    \oiii$\lambda5007$/\hb\ and \hei$\lambda10830$/\pc.
    Figure~\ref{fig:cloudyel} shows results of the dust-free models. The violet red and green lines
    show the observed line ratios of \hei$\lambda10830$/\pc\ and \oiii$\lambda5007$/\hb, which are 
    lower limits of the actual line ratios for the outflow gas (see above). 
    So the region enclosed by the violet red and green lines is the possible parameter space for 
    the outflow gas of this object.
    Model calculations suggest that gas clouds with log~$19 < N_{\rm H}$(cm$^{-2}$) $< 23 $ can generate
    the observed line ratios of this object (Figure~\ref{fig:cloudyel}).
    Both \hei$\lambda10830$/\pc\ and \oiii$\lambda5007$/\hb\ are sensitive to the hydrogen front,
    so the gas which is too thin to contain a hydrogen front or gas which is too thick cannot
    generate the observed line ratios.
    Although the appropriate $N_{\rm H}$ spread 6~dex, the densities $n_{\rm H}$ are confined to 
     4 $\leq$ log~$n_{\rm H}$(cm$^{-3}$) $\leq 6$, which is very close to the condition of absorption line
    outflow gas.
    Thus, we infer that the blueshifted emission lines are produced in the same outflowing material as
    the BALs.
    Based on this assumption, we extract these line ratios from the convergent dusty models
    (see the lower three panels of Figure~\ref{fig:cloudyabs}).
    In Figure~\ref{fig:cloudydustel}, we only shows models with $N_{\rm H} = 10^{22.5}$ cm$^{-2}$ and
    $N_{\rm H} = 10^{23}$ cm$^{-2}$, which cover the parameter space of the convergent dusty models.
    The suitable parameter space to generate the observed line ratios of \thisobj\ is
    $ 3.5 \lesssim$ log $n_{\rm H}(\rm cm^{-3}) \lesssim 5.5$ and $-1.8 \lesssim$ log $U \lesssim -0.7$, 
    which is well consistent with the results of absorption line outflows.
    Combined with the gas parameters determined from the absorption lines, the acceptable parameters  
    (log$n_{\rm H}$, log$N_{\rm H}$, log$U$) of the outflow are $4 < $ log$n_{\rm H}$ $\lesssim 5$,
     $22.5 \lesssim $ log$N_{\rm H}$ $\lesssim 22.9$,
     and $-1.3~\lesssim$~log$U$~$\lesssim~-0.7$.
    Note that the grids with log$n_{\rm H} =4$ cm$^{-3}$ is on the edge of the allowed parameter 
    space.

   %%% 2015July5 night, begins: ------
   \subsubsection{Kinetic luminosity and mass flux of the outflow}

   After analyzing the absorption line and emission line outflows, we find that the hydrogen density 
   of both is quite similar and that the derived values of the ionization parameter $U$ are also consistent.
   This suggests that the blueshifted emission lines are plausibly originated
   in the absorption line outflows.
   If this is the case, we can accurately determine the kinetic properties of
   the outflows by taking advantages of both the absorption lines and the emission
   lines.
   Specifically, the absorption lines, which trace the properties of the outflow
   in the LOS, is good at determining the total column density ($N_{\rm H}$) and
   the velocity (gradient) of the outflow; the emission lines, which trace the
   global properties of the outflows, can determine better the total mass and
   global covering factor of the outflows.
   
\begin{figure}[htbp] 
      \center
     \includegraphics[width=6.4in]{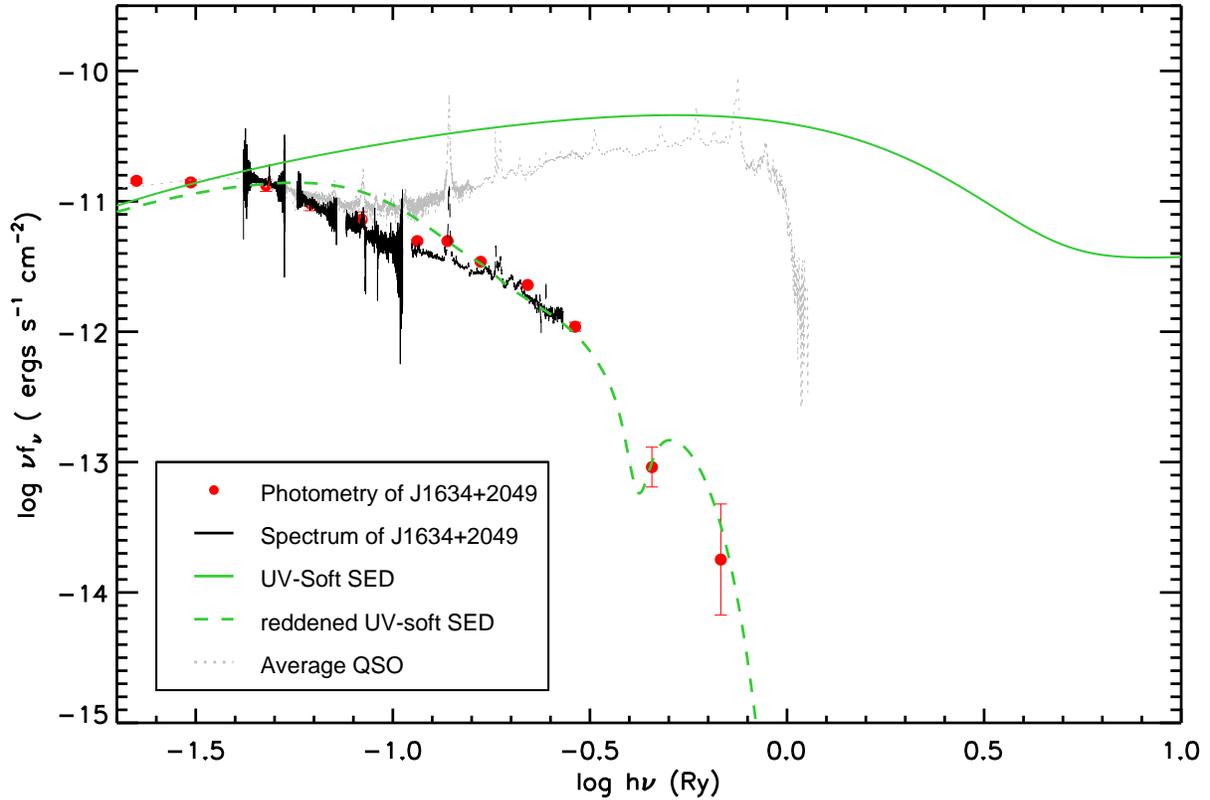}
     \caption{\footnotesize Observed spectrum (the black solid lines) and photometry data 
       (the red dots) of \thisobj. The green solid curve represents the UV-soft SED scaled to 
       the observed flux at the $WISE W1$ band.
       The green dashed curve shows the UV-soft SED reddened with the LMC extinction curve. 
       The gray dotted line is the average QSO spectrum as shown in Figure \ref{fig:sed}.
       \label{fig:photonumber}}
    \end{figure}

   %%% 2015June6--
   As the first step, we determine the distance ($R$) of the outflow 
   (exactly speaking, the part producing the LOS absorption) away from 
   the central source.
   The ionization parameter $U$ depends on $R$ and hydrogen-ionizing photons 
   emitted by the central source ($Q_{\rm H}$),  as follows, 
   \begin{equation}
       U = \frac{Q_{\rm H}}{4\pi R^{2} n_{\rm H} \,c}  ~,
   \end{equation}
   where $n_{\rm H}$ is the density of the outflow and $c$ is the speed of light. 
   The $n_{\rm H}$ has been estimated as log~$n_{\rm H} \approx 4--5$ cm$^{-3}$.
   %% ---xbdong -----
   To determine the $Q_{\rm H}$, we scale the UV-soft SED to the de-reddened flux 
   of \thisobj\ at the $WISE$ $W1$ band ($\sim 3.4 \micron$) (see Figure~\ref{fig:photonumber}),
   and then integrate over the energy range $h\mu \ge 13.6$~eV of this scaled SED.
   This yields $Q_{\rm H} = 4.2 \times 10^{55}$ photons s$^{-1}$.
   To check the reliability of this integration, we integrate also this scaled SED
   over the whole energy range and get $L_{\rm bol} = 4.3 \times 10^{45}$ \ergs,
   which is basically consistent with our estimated \lbol\ independently from the
   obscuration-corrected continuum luminosity at 5100~\AA\ (see \S3.1).
   %%The hydrogen ionizing photon rate are $Q_{\rm H} = 4.2 \times 10^{55}$ photons s$^{-1}$. 
   %% --xbdong end---
   %%
   Using this $Q_{\rm H}$ value together with the derived $n_{\rm H}$ and 
   $U$, the $R$ value can be derived, as listed in Table~\ref{tab:outflowpar}.
   With $R$ being 48--65~pc, the outflow is located exterior to the torus,
   while the extend of the torus is on the scale of
   $\sim 10$~pc \citep{2013A&A...558A.149B}.
   This result is consistent with our qualitative analysis of the normalization
   for the intrinsic spectrum underlying the absorption trough (see \S2.4.3).
  
  %%2015July7 ----
   Assuming that the absorbing material can be described as a thin ($\Delta R/R \ll 1$)
   partially filled shell, the mass-flow rate ($\dot{M}$) and kinetic luminosity ($\dot{E_{k}}$)
   can be derived as follows (see the discussion in \citet{2012ApJ...751..107B}),
   \begin{equation}
     \dot{M} = 4\pi R \Omega\, \mu \,m_{p} N_{H} \, v  ~~,
   \end{equation}
   \begin{equation}
     \dot{E_{k}} = 2\pi R \Omega \, \mu \, m_{p} N_{H} \, v^{3}  ~~,
   \end{equation}
   where $R$ is the distance of the outflow from the central source, $\Omega$ is the global 
   covering fraction of the outflow, $\mu = 1.4$ is the mean atomic mass per proton, 
   $m_{p}$ is the mass of proton,  $N_{\rm H}$ is the total hydrogen column density 
   of the outflow gas, and
   $v$ is the weight-averaged velocity of the absorption trough,  which is directly 
   derived from the trough's profile. 
   The weight-averaged $v$ for the \heiabs$\lambda3889$ absorption trough is $-3837$~\kms.
   %%% 2016Feb27: ---
   %% -- wordy but not to the point: 
   %We note the velocity of absorption-line outflow is much larger than that of 
   %emission-line outflow. This can be explained by the self-shielded effect of the outflow.
   %The absorber, which is the shady side of the outflow, is moving outward along our LOS, 
   %so the velocity of absorption line is close to the real outward velocity of the outflow.
   %The absorber is thick ($N_{\rm H} = 22.5$), and the transmitted continuum is weak 
   %let alone the emission lines.
   %The observed blueshifted emission lines are from a illuminated face that are not shielded 
   %by the absorber, and the outward velocity component along our LOS is much smaller than that 
   %of absorption lines.    
   %Hence it is reasonable to take the velocity of absorption line as the velocity of the outflow.
   %%
   Note that the outflow velocity ($v$) should be calculated with the absorption line velocity,
   not with the outflowing emission line velocity.
  The absorption lines are produced from the absorber moving along our LOS,
  whereas the emission lines originates from gas outflowing along different directions
  with respect to the observer.
  Thus the observed outflow velocity of an  emission line
  is a sum of the projected velocities of the outflowing gas along different directions,
  and should be smaller than the outflow velocity of the absorbing material; this is just as we observed.

   %% paragraph \omega --- :
   We estimated the global covering fraction ($\Omega$) for \thisobj\ by comparing the 
   measured EW(\oiii 5007) with the predicted one by the best {\it Cloudy} model (see \S3.4.2).
   Although to this end, theoretically it is better to use recombination lines such
   as \hb\ and \hei$\lambda10830$; there is, however, no good measurements of the \hb\
   for the relatively broad, blueshifted component (see discussion in \S\ref{subsec:cloudyel})
   and that component of \hei$\lambda10830$ is heavily affected by the absorption trough.
   The EW(\oiii) value is affected by the dust extinction both to the continuum and
   the \oiii~$\lambda$5007 emission.
   We make simple and reasonable assumptions as follows (cf. \S2.4.2): the continuum suffers
   dust extinction to the same degree of the broad lines with E$_{\rm B-V} = 0.64$, and the
   \oiii~$\lambda$5007 emission, within the range of dust-free and  the broad-line one.
   After corrected for the dust extinction, the actual EW(\oiii) should be in the range 
   of 4.3--24.8~\AA.
   %is important to calculate 
   %Assuming emitting region of [OIII]$\lambda$5007 outflow is dust free, while continuum 
   %experiences the same extinction with BLR, which are de-reddened with LMC extinction by 
   %E(B-V) = 0.64. In this case, the measured EW(\oiii) is 24.8 \AA.
   %Assuming the extinction of [OIII]$\lambda$5007 outflow is the same with that of continuum, 
   %the measured EW(\oiii) is 4.3 \AA. 
   %The two values are the possible range of measured EW(\oiii).
   %% ---the order of 4.3 and 24.8 is wrong. - 2015July7, corrected. ---
   %
   Here the outflowing \oiii\ flux is the same as used in \S\ref{subsec:cloudyel} 
   and the continuum flux is measured from the decomposed power-law component at 5007\AA.
   In {\it Cloudy} modeling, the emergent \oiii\ flux is output with the 
   covering fraction being assumed to be 1.
   The derived EW(\oiii) is 82~\AA\ for the model with 
   $n = 10^{4.5}$ cm$^{-3}$, and 142~\AA\ for $n = 10^{5}$ cm$^{-3}$. 
   Thus, the global cover fraction ($\Omega$) for \thisobj\  
   is estimated to be in the range of
   5.2--30.1\% for $n = 10^{4.5}$ cm$^{-3}$, 
   or 3.0--17.4\% for $n = 10^{5}$ cm$^{-3}$.
   Likewise, we estimate the $\Omega$ for the outflow emitting \hei$\lambda10830$, 
   yielding  72.9--100\% (the case of $n = 10^{4.5}$ cm$^{-3}$)
   or 43.7--71.7\% ($n = 10^{5}$ cm$^{-3}$), 
   which are much lager than those for \oiii~$\lambda$5007.
   The large difference between the $\Omega$ values estimated based on
   \hei$\lambda10830$ and \oiii\
   may be mainly due to the measurement uncertainty of the outflowing component of
   \hei$\lambda10830$, and/or may reflect the inhomogeneity of the outflowing gas.
   \oiii\ is a forbidden line that traces the region of low density only and
   \hei$\lambda10830$, a recombination line, can be generated in much broader spatial regions.
   Here we conservatively  adopt the $\Omega$ value estimated from \oiii.
   Thus, the kinetic luminosity and mass loss rate are calculated as summarized in 
   Table~\ref{tab:outflowpar}.

 %%2015July7, night ----
\section{Discussion and Summary}
% \section{Summary}
 
   %% overview---
   We performed a comprehensive multiwavelength study of the properties of \thisobj,  
   a local quasar with high IR luminosity of \rev{$L_{\rm IR} = 10^{11.91} \lsun$} and 
   signatures of outflows both in BALs and blueshifted emission lines.
   %% SFR ---
   The high IR luminosity indicates recent violent star formation activities with 
   SFR = 140 $\pm 43~\msun~\rm~yr^{-1}$, which is also indicated by the PAH emission lines
   with a derived $SFR = 141~\pm 71 \msun \rm yr^{-1}$.
   %% Interestingly, it's dull in the optical, so the SF is obscured---
   Interestingly, in the UV and optical bands there are few signs of star 
   formation activities, and the decomposition of the SDSS images 
   demonstrate the host galaxy is of early type in terms of its structural 
   property (e.g., S\'{e}rsic $n=4$) and the colors (e.g., $g-r = 0.93$).
   %The analysis on the continuum and narrow emission lines show that the star formation region is 
   %heavily obscured by dust.
   %In NIR band, we can roughly estimate the contribution of star formation from the strong \pa\ 
   %narrow emission line. --- seems not need to say these two sentences.
   %
   This quasar has a circumgalactic ring on scales of $\sim30$~kpc.
   The ring has almost the same colors as the host galaxy 
   (namely the S\'{e}rsic $n=4$ component). 
   Yet it is unclear at this point whether or not the galactic ring has a similar 
   intense dust-enshrouded star formation.
   There are two small galaxies to the west of it within 20\arcsec\ on the sky, 
   the three of them being in an almost a linear configuration.
   We spectroscopically observed the galaxies and obtained the redshift for the 
   relatively bigger one (C1; the closer to \thisobj), $z= 0.1298^{+0.007}_{-0.005}$.
  % The redshift difference between \thisobj\ and C1 means a line-of-sight velocity 
  % difference of 451~\kms\ only.
  % We can reasonably suggest that the galactic ring is produced by the head-on 
  % collision between the nearby galaxy/galaxies and \thisobj; 
  % if so, the collision would occur less than $\sim 160$~Myr ago.  

   In the optical and near-infrared spectra, there are several strong BALs and 
   blueshifted emission lines in addition to the normal, 
   BLR- and NLR-originated emission lines, suggestive of the AGN outflows.
   These lines can be used to derive (mutually constrain) the physical properties 
   of the outflowing gas by confronting the observed with the modeling results of 
   the photoionization software {\it Cloudy}.
   The appearance of the BAL of neutral Sodium, which is rare in 
   an AGN environment, suggests that the outflowing gas is thick and dusty.
   %%We determine the physical properties for both of them using the photoionization models.
   The physical parameters we determined with {\it Cloudy} for absorption line and 
   emission line outflows are very close, with $10^{4.5} \leq n_{\rm H} \leq 10^5$ cm$^{-3}$, 
   $10^{-1.3} \leq U \leq 10^{-1.0}$ and $N_{\rm H} \approx 10^{22.5}$ cm$^{-2}$.
   This similarity  %%--physical properties of the absorption line and emission line outflows 
   suggests that those absorption  and emission lines should be generated in the 
   common outflowing gas.
   Using the absorption lines to derive the total column density of the outflow and 
   the emission lines to obtain the global covering factor, we estimate the distance 
   of the absorbing material to the central source, $R \sim$48--65~pc, exterior to the torus.
   These derived parameters allow us to calculate the kinetic luminosity and mass 
   loss rate of the outflow.
   %%%
   The results listed in Table~\ref{tab:outflowpar} indicate that the outflow 
   observed in \thisobj\ processes a large kinetic luminosity, 
   which is high enough to play a major role in the AGN feedback.
   Previous studies suggest that AGN feedback typically requires a mechanical energy input of 
   roughly 0.5--5\% of Eddington luminosity of the quasar to heat the cold gas and 
   quench the star formation activities in the host galaxy  
   (e.g., Hopkins \& Elvis 2010; Scannapieco \& Oh 2004).
   %The dynamical time of the outflow  $t_{d} \equiv R/v \approx $12,000--16,500~yrs,
   %which is much smaller than the elapsed time after the galactic collision.
   % So the outflow must not be high-velocity gas streams caused by the galactic collision,
   % but be of an AGN origin.

   %%the last paragraph---
   Taking the multiwavelength results altogether, we can infer the whole story as follows. 
   \thisobj\ could have been a disk galaxy with abundant gas, and was collided through
   by one or two small galaxies with masses of $\sim 1\times 10^{10} \msun$. 
   The violent head-on collision destroyed the disk and formed a circumgalactic ring.
   The collision also triggered violent star formation in the host galaxy,
   with the star formation regions being heavily reddened by
   dust now and betraying itself in the IR bands only.
   We may infer that the violent star formation is in the circumnuclear region,
   as the optical images and spectrum demonstrate that the large scale of the host
   galaxy is an elliptical galaxy dominant by an old stellar population.
   Indeed, in the generally believed scenario of galaxy merger/collision and AGN feedback,
   as described in the Introduction section, the large-scale cold gas of the interacting
   galaxies loses angular momentum and is driven into the nuclear region due to the
   gravitational potential well; the infall of the cold gas triggers nuclear star
   formation and also feeds the central SMBH.
   This inference is further supported by the fact that the narrow emission lines 
   (dominated by the AGN according to the BPT diagrams) as well as the broad emission 
   lines are seriously obscured by dust, hinting at a common origin of the dust to the 
   optical emission of both the AGN and star formation.
   The existence of neutral sodium BAL also indicates that the nuclear
   region is dusty. 
   With huge kinetic luminosity, the outflows launched by the AGN in turn will 
   blow away the gas and dust around the nucleus and even inhibit the star formation 
   in the host galaxy soon or later, as the co-evolution scenario suggests.
   At this time point, \thisobj\ is just like the prototypal QSO/LIRG composite object, 
   Mrk~231, in the transitional phase  emerging out of the dust. 
   According to the differences in both the degree of dust extinction and 
   the mid- and far-infrared SED shape between \thisobj\ and Mrk~231,
   \thisobj\ should be at the phase immediately after Mrk~231 in the co-evolutionary sequence.
   So it is a rare object for us to detailedly and quantitatively study (or test) 
   the co-evolution scenario of galaxy and SMBH.
   In particular, taking the advantage of its nearness,
   we will carry out spatially resolved observations of the cold gas distribution
   and kinematics on the circumgalactic scale (by, e.g., JVLA), or even on the scale
   within the host galaxy (by, e.g., ALMA), to directly investigate the cold gas---the vital
   starring actor playing in the (co-)evolutionary scenario.\\

%%%%%%%%%%%%%%%
%\\ acknowledge
   We thank Xiao-Bo~Dong for comments on the manuscript and improving the English and
   Lei~Hao for the helpful discussions on the MIR properties of \thisobj.
   We thank Baoli~Lun for the help of the spectroscopic observing by the Lijiang 2.4m telescope.
   W.-J.~Liu particularly thanks Shaohua Zhang, Zhenzhen Li and Xiang~Pan for the helpful
   discussions on the {\it Cloudy} models and thanks Ting~Xiao for the helpful discussions on the
   neighboring galaxies and the galactic ring.
   This work is supported by the SOC program (CHINARE2012-02-03),
   the Natural Science Foundation of China grants (NSFC 11473025, 11033007, 11421303),
   National Basic Research Program of China (the 973 Program 2013CB834905),
   and the funding from the Key Laboratory for the Structure and Evolution of Celestial Objects, 
   Chinese Academy of Sciences (OP201408).
   W.-J. L. acknowledges support from the Natural Science Foundation of China grants (NSFC 11203021, 11573024).
   X. W. acknowledges support from the Natural Science Foundation of China grants (NSFC 11503025) and
   the Fundamental Research Funds for the Central Universities; 
   X. S. acknowledges support from the Natural Science Foundation of China grants (NSFC 11573001, NSFC 11233002); 
   J.-G. W. acknowledges support from the Natural Science Foundation of China grants (NSFC 11303085).
   This work has made use of the data products of the SDSS, data obtained by the
   Lijiang 2.4m telescope and through the Telescope Access Program (TAP) in 2012B (PI: Xinwen~Shu),
   2014A (PI: Tuo~Ji).
   We acknowledge the support of the staff of the Lijiang 2.4m telescope.
   Funding for the telescope has been provided by CAS and the People's Government of
   Yunnan Province.
   TAP is funded by the Strategic Priority Research Program ``The Emergence of 
   Cosmological Structures'' (grant no. XDB09000000), the National Astronomical Observatories, 
   the Chinese Academy of Sciences, and the Special Fund for Astronomy from the Ministry of 
   Finance.
   Observations obtained with the Hale Telescope at Palomar Observatory were obtained 
   as part of an agreement between the National Astronomical Observatories, the Chinese 
   Academy of Sciences, and the California Institute of Technology.

%%%% Figure 1

%%% Figure 2

  %%% Figure 3 below----

 %%%%%%Figure 4

%%%% Figure 5
% \begin{figure}[tbp]
%   \center
%   \includegraphics[width=5.1in]{fig_5a.eps}\includegraphics[width=1.3in]{fig_5b.eps}
%   \includegraphics[width=5.1in]{fig_5c.eps}\includegraphics[width=1.3in]{fig_5d.eps}
%   \caption{Different models to fit the \pa\ (top panels) and \ha\ (bottom panels)
%            emission-line regions.
%            Panels a, b, c, and d show the fitting results of fitting \pa\ and \ha\
%        separately using different models (see the text);
%        panels e1 (\pa) and e2 (\ha) show the fitting results under the assumption
%        that the narrow-line and broad-line \ha\ components have the same profiles
%        with those of \pa\ respectively.
%        Panel d1 and panel d2 are the best-fit models that we last adopt for \pa\
%        and \ha. \label{fig:hapafit}}
% \end{figure}

%%%%% Figure 5

 %%%% Figure 6

%%% Figure 7

%% Figure 8 ---

%%% Figure 9

 %%% Figure 10

   %%% Figure 11

   %%%%%%%% Figure 12

  %% Figure 13----

    \clearpage
%---------------------------------------------------------------------------------

   %%%% -Table 1 - Photometric Data -----------
   \begin{deluxetable}{lrrll}
    \tablecolumns{5}
    \tablewidth{0pc}
    \tabletypesize{\footnotesize}
    \tablecaption{Photometric Data \label{tab:photometry}}
   
    \tablehead{
      \colhead{Band} & \colhead{Mag/Flux} &
      \colhead{Facility} & \colhead{Obs. Date} &  \colhead{Reference} \\
      \colhead{}     & \colhead{}          & 
      \colhead{}     & \colhead{(UT)} & \colhead{}}
    \startdata
     FUV                                & 21.75$\pm$0.46  mag& $GALEX$  & 2006-12-23 & 1 \\     
     NUV                                & 20.47$\pm$0.17  mag& $GALEX$  & 2006-12-23 & 1 \\
     Petrosian $u$                      & 18.94$\pm$0.04  mag& SDSS     & 2003-06-23 & 2, 3 \\
     Petrosian $g$                      & 17.76$\pm$0.01  mag& SDSS     & 2003-06-23 & 2, 3\\
     Petrosian $r$                      & 16.95$\pm$0.01  mag& SDSS     & 2003-06-23 & 2, 3\\
     Petrosian $i$                      & 16.30$\pm$0.01  mag& SDSS     & 2003-06-23 & 2, 3\\
     Petrosian $z$                      & 16.07$\pm$0.01  mag& SDSS     & 2003-06-23 & 2, 3\\
     $J$                                & 14.65$\pm$0.04  mag& 2MASS    & 1997-06-09 & 4 \\ 
     $H$                                & 13.45$\pm$0.03  mag& 2MASS    & 1997-06-09 & 4 \\ 
     $K_{s}$                            & 12.25$\pm$0.03  mag& 2MASS    & 1997-06-09 & 4 \\ 
     $W1$                               & 10.73$\pm$0.02  mag& $WISE$   & 2010-05-29 & 5 \\
     $W2$                               &  9.72$\pm$0.02  mag& $WISE$   & 2010-05-29 & 5 \\
     $W3$                               &  7.07$\pm$0.02  mag& $WISE$   & 2010-02-21 & 5 \\
     $W4$                               &  4.61$\pm$0.02  mag& $WISE$   & 2010-02-21 & 5 \\
     IRAC 8$\micron$\tablenotemark{a}   & 0.035$\pm$0.001  Jy& $Spitzer$& 2008-04-30 & 6 \\
     $IRAS$ 12$\micron$                 & 0.085$\pm$0.019  Jy& $IRAS$   & 1991-05-30 & 7 \\
     IRAC 16$\micron$\tablenotemark{a}  & 0.067$\pm$0.002  Jy& $Spitzer$& 2008-04-30 & 6 \\ 
     IRS  22$\micron$\tablenotemark{a}  & 0.110$\pm$0.003  Jy& $Spitzer$& 2008-04-30 & 6 \\
     MIPS 24$\micron$\tablenotemark{a}  & 0.114$\pm$0.003  Jy& $Spitzer$& 2008-04-30 & 6 \\
     $IRAS$ 25$\micron$                 & 0.141$\pm$0.018  Jy& $IRAS$   & 1991-06-06 & 7 \\
     $IRAS$ 60$\micron$                 & 0.559$\pm$0.045  Jy& $IRAS$   & 1991-03-08 & 7 \\
     $AKARI$ 65$\micron$                & 0.239            Jy& $AKARI$  & 2011-09-08 & 8 \\
     $AKARI$ 90$\micron$                & 0.579$\pm$0.060  Jy& $AKARI$  & 2011-09-11 & 8  \\ 
     $IRAS$ 100$\micron$                & 1.172$\pm$0.199  Jy& $IRAS$   & 1991-04-11 & 7 \\
     $AKARI$ 140$\micron$               & 1.468$\pm$1.753  Jy& $AKARI$  & 2011-09-11 & 8  \\
     1.4 GHz                            & 21.97$\pm$0.147 mJy& FIRST    & 1998-10-07 & 9 %%\\
    \enddata
    \tablerefs{(1)\citet{2007ApJS..173..682M} (2) \citet{2000AJ....120.1579Y}
    (3)\citet{2009ApJS..182..543A} (4) \citet{2006AJ....131.1163S}
    (5)\citet{2010AJ....140.1868W} (6) \citet{2004ApJS..154...18H}
    (7)\citet{1992ifss.book.....M} (8) \citet{2009sitc.conf.4018D}
    (9) \citet{1994ASPC...61..165B}}
    \tablenotetext{a}{Synthetic Photometry data from $Spitzer$ IRS spectrum given by $Spitzer$ 
                      data archive.}
    \end{deluxetable}

%%---- Table 2
\begin{deluxetable}{llcccccc}
  \tablecolumns{8}
  \tablewidth{0pc}
  \tabletypesize{\footnotesize}
  \tablecaption{GALFIT Decomposition \label{tab:galfit}}
  \tablehead{\colhead{Band} & \colhead{Component \tablenotemark{a}} & \colhead{$m$ \tablenotemark{b}} &
             \colhead{$M$ \tablenotemark{c}} & \colhead{$m$ (in 3\arcsec)\tablenotemark{d}}  & 
	     \colhead{$M$ (in 3\arcsec)\tablenotemark{e}} & \colhead{$n$\tablenotemark{f}}  & 
	     \colhead{$r$(\arcsec/kpc)\tablenotemark{g}}  }
  \startdata
  SDSS $g$  & PSF         & 18.68  & -20.49    & 18.74  & -20.43   &    &            \\
            & S\'{e}rsic  & 17.95  & -21.22    & 19.13  & -20.04   & 4  & 2.41/5.56  \\
      	    & Ring        & 19.85  & -19.32    &        &          &    &            \\
	    & C1          & 19.98  & -19.15    &        &          &3.32& 0.55/1.26  \\
  SDSS $r$  & PSF         & 18.16  & -20.93    & 18.24  & -20.85   &    &            \\
            & S\'{e}rsic  & 17.02  & -22.07    & 17.99  & -22.07   & 4  & 1.82/4.20  \\
  	    & Ring        & 18.91  & -20.18    &        &          &    &            \\
	    & C1          & 18.95  & -20.10    &        &          &2.46& 0.49/1.13  \\
  SDSS $i$  & PSF         & 17.51  & -21.53    & 17.56  & -21.48   &    &            \\
            & S\'{e}rsic  & 16.52  & -22.52    & 17.29  & -21.75   & 4  & 1.21/2.80  \\
  	    & Ring        & 18.43  & -20.61    &        &          &    &            \\
	    & C1          & 18.61  & -20.40    &        &          &2.34& 0.45/1.03  %%\\
  \enddata
  \tablenotetext{a}{Components used in the fitting schemes.}
  \tablenotetext{b}{The integrated magnitudes on the Vega system, not corrected for Galactic extinction.
                    The ring magnitude is derived from the residual images.}
  \tablenotetext{c}{The absolute magnitude after Galactic extinction correction.}
  \tablenotetext{d}{The integrated magnitudes on the Vega system within 3\arcsec\ diameter aperture which 
                    corresponding to the fiber aperture of SDSS spectrum, not corrected for Galactic extinction.}
  \tablenotetext{e}{The absolute magnitude within 3\arcsec\ after Galactic extinction correction.}
  \tablenotetext{f}{The S\'{e}rsic index.}
  \tablenotetext{g}{The effective radius of the S\'{e}rsic component in unit of arcseconds and the corresponding 
                    scale length in unit of kpc.}
  \end{deluxetable}

 %%% Table 3 ------------------------%%
\begin{deluxetable}{llcc}
\tablecolumns{4}
\tablewidth{0pc}
\tabletypesize{\scriptsize}
\tablecaption{Emission Line Parameters \label{tab:emline}}
\tablehead{
  \colhead{Emission Line} & \colhead{Centroid\,\tablenotemark{a}} & \colhead{FWHM\,\tablenotemark{b}} & 
  \colhead{Flux}  \\
  \colhead{ }             & \colhead{(\AA)}    & \colhead{(\kms)} & 
  \colhead{(10$^{-17}$ erg~s$^{-1}$~cm$^{-2}$)}}
\startdata
 \ [O\,II]$\lambda3727$(narrow)\           & 3727.39$\pm$0.21  & 491$\pm$5   &  184$\pm$3    \\
 \ [O\,II]$\lambda3727$(outflow)\          & 3720.19$\pm$2.22  & 738$\pm$5   &   89$\pm$8    \\
 \ H$\gamma$(broad)\,\tablenotemark{c} \     & 4342.43           & 2955        & 487$\pm$25    \\
 \ H$\gamma$(narrow)\,\tablenotemark{d}\     & 4340.74           & 343         &  8$\pm$5      \\
 \ H$\gamma$(outflow)\,\tablenotemark{e}\     & 4327.58           & 737   &  7$\pm$9      \\
 \ H$\beta$(broad)\,\tablenotemark{c}\     & 4863.52           & 2955        &  1378$\pm$14  \\
 \ H$\beta$(narrow)\,\tablenotemark{d}\     & 4861.63           & 343         &  42$\pm$3     \\
 \ H$\beta$(outflow)  \                    & 4846.50           & 737$\pm$20       &  58$\pm$5     \\
 \ [O\,III]$\lambda5007$ \                 & 4992.80$\pm$0.32  & 1707$\pm$18 &  1043$\pm$12  \\
 \ [O\,I]$\lambda6300$\,\tablenotemark{f}\   & 6299.87$\pm$0.53  & 421         &  61$\pm$6     \\
 \ H$\alpha$(broad)   \                    & 6565.74$\pm$0.18  & 2955$\pm$12 &  8719$\pm$66  \\
 \ H$\alpha$(narrow)  \                    & 6563.18$\pm$0.03  & 343$\pm$5   &  427$\pm$8    \\
 \ H$\alpha$(outflow)\,\tablenotemark{e}\   & 6543.30$\pm$0.13  & 737  &  351$\pm$17   \\
 \ [N\,II]$\lambda6583$\                   & 6583.05$\pm$0.03  & 370$\pm$4   &  458$\pm$6    \\
 \ [S\,II]$\lambda6716$ \                  & 6716.42$\pm$0.12  & 421$\pm$10  &  148$\pm$5    \\
 \ [S\,II]$\lambda6731$\,\tablenotemark{f}\                & 6730.80           & 421         &  106$\pm$4    \\
 \ \hei$\lambda$10830(broad)\,\tablenotemark{g}\    & 10840.33$\pm$1.48 & 3803 &  994$\pm$53   \\
 \ \hei$\lambda$10830(narrow) \             & 10831.33$\pm$0.45 & 178$\pm$12  &  162$\pm$25   \\
 \ \hei$\lambda$10830(outflow$^{\rm N}$)\       & 10801.14$\pm$0.33 & 140$\pm$23  &  21$\pm$7     \\
 \ \hei$\lambda$10830(outflow$^{\rm B}$)\       & 10782.34$\pm$0.33 & 1612$\pm$23 &  764$\pm$37   \\
 \ P$\gamma$(broad)\,\tablenotemark{g}\        & 10948.39          & 3803        &  874$\pm$34   \\
 \ P$\gamma$(narrow)                 \     & 10940.50$\pm$0.31 & 204$\pm$18  &  69$\pm$10    \\
 \ P$\gamma$(outflow)\,\tablenotemark{h}\     & 10907.00          & 575         &  50$\pm$11    \\
 \ P$\alpha$(broad)                  \     & 18768.64$\pm$0.20 & 3803$\pm$103&  2484$\pm$79  \\
 \ P$\alpha$(narrow)                 \     & 18755.96$\pm$0.19 & 310$\pm$11  &  486$\pm$13   \\
 \ P$\alpha$(outflow)                \     & 18697.69$\pm$1.64 & 575$\pm$75  &  146$\pm$16   %%\\
\enddata
\tablenotetext{a}{Vacuum rest frame wavelength.}
\tablenotetext{b}{Corrected for instrumental broadening.}
\tablenotetext{c}{Adopting the profile of the broad component of \ha.}
\tablenotetext{d}{Adopting the profile of the narrow component of \ha.}
\tablenotetext{e}{Adopting the profile of the outflow component of \hb.}
\tablenotetext{f}{Adopting the profile of [S~II]$\lambda6716$.}
\tablenotetext{g}{Adopting the profile of the broad component of \pa.}
\tablenotetext{h}{Adopting the profile of the outflow component of \pa.}
\end{deluxetable}

%%%%------- Table 4
\begin{deluxetable}{llll}
   \tablecolumns{4}
   \tablewidth{0pt}
   \tabletypesize{\small}
   \tablecaption{Balmer Decrement \label{tab:decrement}}
   \tablehead{\colhead{Decrement} & \colhead{Component} & \colhead{Value}     & \colhead{$E_{\rm B-V}$} }
   \startdata
     H$\alpha$/H$\beta$\tablenotemark{a}  & broad   & \rev{6.33$\pm$0.08}  & 0.64$\pm$0.01 \\
     H$\alpha$/H$\beta$\tablenotemark{b}  & narrow  & 10.17$\pm$0.75  & 1.05$\pm$0.06 \\
     H$\alpha$/H$\beta$\tablenotemark{b}  & outflow (narrow) &  6.05$\pm$0.60  & 0.60$\pm$0.09 \\
     P$\alpha$/H$\beta$\tablenotemark{c}  & broad   &  1.80$\pm$0.06  & 0.66$\pm$0.01 \\
     P$\alpha$/H$\beta$\tablenotemark{c}  & narrow  & 11.57$\pm$0.90  & 1.39$\pm$0.03 \\
     P$\alpha$/H$\beta$\tablenotemark{c}  & outflow (narrow) &  2.52$\pm$0.35  & 0.79$\pm$0.05 %%\\
     %P$\alpha$/H$\alpha$                  & broad   &  0.28$\pm$0.01  & 0.62$\pm$0.02 \\
     %P$\alpha$/H$\alpha$                  & narrow  &  1.14$\pm$0.04  & 1.53$\pm$0.02 \\
     %P$\alpha$/H$\alpha$                  & outflow &  0.42$\pm$0.05  & 0.87$\pm$0.08 \\
   \enddata
   \tablenotetext{a}{Intrinsic \ha/\hb\ for BLR is 3.06 \citep{2008MNRAS.383..581D}.}
   \tablenotetext{b}{Intrinsic \ha/\hb\ for NLR and outflowing gas is 3.1 \citep{2008MNRAS.383..581D}.}
   \tablenotetext{c}{Intrinsic \pa/\hb\ is 0.34 \citep{1984PASP...96..393G}.}
   \end{deluxetable}

%% Table 5 --
\begin{deluxetable}{lcr}
 \tablecolumns{3}
 \tablewidth{0pc}
 \tabletypesize{\footnotesize}
 \tablecaption{Recovered Dust Features and Line Strengths \label{tab:mirpar}}
 \tablehead{\colhead{Emission Line} & \colhead{Intensity\,\tablenotemark{a}}                & \colhead{$EW$\,\tablenotemark{b}}  \\
            \colhead{}              & \colhead{(10$^{-20}$ W~cm$^{-2}$)} & \colhead{($\micron$)}
	   }
 \startdata
 \ PAH 6.2$\micron$            \  & 1.47$\pm$0.09  &  \rev{0.050$\pm$ 0.003}  \\
 \ PAH 7.7$\micron$ complex    \  & 3.06$\pm$0.16  &  \rev{0.135$\pm$ 0.007}  \\
 \ PAH 8.3$\micron$            \  & 0.57$\pm$0.12  &  \rev{0.028$\pm$ 0.006}  \\
 \ PAH 8.6$\micron$            \  & 1.08$\pm$0.08  &  \rev{0.055$\pm$ 0.004}  \\
 \ PAH 11.3$\micron$ complex   \  & 1.25$\pm$0.08  &  \rev{0.085$\pm$ 0.005}  \\
 \ PAH 12.0$\micron$           \  & 0.10$\pm$0.09  &  \rev{0.007$\pm$ 0.006}  \\
 \ PAH 12.6$\micron$ complex   \  & 1.44$\pm$0.14  &  \rev{0.106$\pm$ 0.010}  \\
 \ PAH 13.6$\micron$           \  & 0.43$\pm$0.10  &  \rev{0.033$\pm$ 0.008}  \\
 \ PAH 16.4$\micron$           \  & 0.13$\pm$0.07  &  \rev{0.011$\pm$ 0.006}  \\
 \ PAH 17$\micron$ complex     \  & 0.34$\pm$0.25  &  \rev{0.029$\pm$ 0.019}  \\
 \ [NeII]$\lambda12.8\micron$  \  & 0.32$\pm$0.03  &  \rev{0.034$\pm$ 0.004}  \\
 \ [NeIII]$\lambda15.6\micron$ \  & 0.12$\pm$0.03  &  \rev{0.014$\pm$ 0.004}  %%\\
\enddata
%%2015Feb26:---
\tablenotetext{a}{Uncertainties are given by the PAHFIT code; see \citep{2007ApJ...656..770S} for the detail.}
\tablenotetext{b}{Uncertainties are estimated according to error propagation formula.}

\end{deluxetable}

%% Table 6 ----   
\begin{deluxetable}{llllllllll}
\tablecolumns{10}
\tablewidth{0pc}
\tabletypesize{\footnotesize}
\tablecaption{Physical Properties of the Outflow \label{tab:outflowpar}}
\tablehead{\colhead{log$n_{\rm H}$} & \colhead{log$N_{\rm H}$} & \colhead{log$U$} & 
\colhead{$v$}        & \colhead{$R$}            & \colhead{$\Omega$}       & \colhead{log$L_{\rm Bol}$\tablenotemark{a}} &
	   \colhead{$\dot{M}$}  & \colhead{log$\dot{E}_{\rm k}$} &  \colhead{$\dot{E}_{k}/L_{\rm Bol}$} \\
	   \colhead{(cm$^{-3}$)}    & \colhead{ (cm$^{-2}$) }  & \colhead{  } &
	   \colhead{ \kms }     & \colhead{(pc)}           & \colhead{ (\%) }         & \colhead{(ergs s$^{-1}$)} &
	   \colhead{(\msun yr$^{-1}$)} & \colhead{(ergs s$^{-1}$)} & \colhead{(\%)}}
\startdata
  4.0 \tablenotemark{b}& 22.85 & -0.55 & -3837 & 64.1 & 18.1 - 100 & 45.6 & 455--2513 & 45.3--46.0 & 49.2--272  \\
  4.5 & 22.6 & -1.0 & -3837 & 60.6 &  5.2 -- 30.1  & 45.6 & 70--402  &44.5--45.3 & 7.5--43.5  \\
  5.0 & 22.5 & -1.3 & -3837 & 48.1 &  3.0 -- 17.4  & 45.6 & 25--147  &44.1--44.8 & 2.8--15.9  %%\\
\enddata
\tablenotetext{a}{\rev{The $L_{\rm bol}$ here is calculated from UV-soft SED used in CLOUDY models, which corresponds to the 
estimated hydrogen-ionizing photons ($Q_{\rm H}$) used to estimated the distance of outflows. 
The $L_{\rm bol}$ in this table is basically consistent with the $L_{\rm bol}$ calculated from $\lambda L_{\lambda}(5100\AA)$.}}
\tablenotetext{b}{Grid with $n_{\rm H}$ = 10$^{4}$ cm$^{-3}$ is on the edge of the acceptable parameter space.}
\end{deluxetable}

\end{document}